\documentclass[conference,review=true,anonymous=true]{IEEEtran} 
\IEEEoverridecommandlockouts

\usepackage{etoolbox}
\usepackage{ifthen}

\newboolean{makeXTRA}
\setboolean{makeXTRA}{false} 

\makeatletter%
\newboolean{maketitleAfterAllMeta}
\@ifclassloaded{acmart}%
{\setboolean{maketitleAfterAllMeta}{true}}%
{\setboolean{maketitleAfterAllMeta}{false}}%
\makeatother%

\usepackage[utf8]{inputenc}

\usepackage[T1]{fontenc}

\usepackage{microtype}

\usepackage{anyfontsize}
\usepackage[10pt]{moresize}

\usepackage{amsmath}
\usepackage{amsfonts}

\makeatletter%
\@ifclassloaded{acmart}{}{\usepackage{amssymb}}
\makeatother%

\usepackage{siunitx}
\usepackage{xfrac}

\usepackage{xspace}

\usepackage[normalem]{ulem}

\makeatletter%
\@ifclassloaded{acmart}{}%
{\usepackage[usenames,dvipsnames,table]{xcolor}}
\makeatother%

\makeatletter%
\@ifclassloaded{acmart}{}
{\@ifclassloaded{elsarticle}{}
{\usepackage{cite}}}
\makeatother%

\usepackage{setspace}

\usepackage{verbatim}

\usepackage{booktabs}
\usepackage{makecell}
\usepackage{multicol}
\usepackage{multirow}
\usepackage{tabularx}
\usepackage{array}

\usepackage{stfloats}

\makeatletter
\@ifclassloaded{IEEEtran}{%
  \let\MYcaption\@makecaption%
  \usepackage[font=footnotesize]{subcaption}%
  \let\@makecaption\MYcaption%
}{%
  \usepackage{subcaption}%
}%
\makeatother

\usepackage[textsize=tiny]{todonotes}

\makeatletter%
\@ifclassloaded{IEEEtran}{}{}%
\usepackage{enumitem}
\makeatother%

\usepackage{quoting}

\usepackage{listings}

\lstdefinestyle{mystyle}{
    language=Python,           
    basicstyle=\ttfamily,      
    keywordstyle=\color{blue}, 
    frame=single,              
    breaklines=true,           
    captionpos=b               
}

\makeatletter%
\@ifclassloaded{acmart}%
{}{}%
\@ifclassloaded{elsarticle}%
{}{}
\@ifclassloaded{llncs}%
{}{}
\@ifclassloaded{jpconf}%
{%
  \usepackage{doi}%
}{}
\@ifclassloaded{IEEEtran}%
{%
  
  \newcommand{\ifIEEEtran}[1]{#1}
}%
{%
  \newcommand{\ifIEEEtran}[1]{}
  \usepackage{flushend}
}%
\makeatother%

\makeatletter%
\@ifpackageloaded{hyperref}%
{}%
{%
  \usepackage[hyphens]{url}
  \usepackage[breaklinks=true,bookmarks=false]{hyperref}
  \hypersetup{
    colorlinks=false,%
    pdfborder = 0 0 0
  }
}%
\makeatother%

\usepackage[capitalize,noabbrev,capitalize]{cleveref} 

\usepackage{stackengine}
\usepackage{graphicx}
\usepackage{tikz}
\usepackage{pgfplots}
\usepackage{pgfplotstable}

\ifthenelse{\boolean{makeXTRA}}
  {\newcommand{\XTRA}[1]{\phantom{}\begingroup\slshape\color{RoyalBlue}\ignorespaces#1\ignorespaces\endgroup}}
  {\newcommand{\XTRA}[1]{}}

\DeclareMathAlphabet{\mathpzc}{OT1}{pzc}{m}{it}

\makeatletter
\newcommand*{\textoverline}[1]{$\overline{\hbox{#1}}\m@th$}
\makeatother

\makeatletter
\@ifclassloaded{IEEEtran}{%
}{%
  
}%

\makeatletter 
\newcommand{\linebreakand}{%
  \end{@IEEEauthorhalign}
  \hfill\mbox{}\par
  \mbox{}\hfill\begin{@IEEEauthorhalign}
}
\makeatother 

\newlist{myReuseInterval}{enumerate}{1}
\setlist[myReuseInterval]{label=(R$_{\arabic*}$),nosep}

\crefname{section}{\S}{\S}
\crefformat{section}{#2\S#1#3}

\crefname{figure}{Fig.}{Figs.}
\Crefname{figure}{Figure}{Figures}

\crefname{equation}{Eq.}{Eqs.}

\definecolor{shadecolor}{gray}{0.9} 

\makeatletter
\@ifpackageloaded{algorithm2e}{\@ifclassloaded{IEEEtran}{%
  \SetAlFnt{\small}
  \SetAlCapFnt{\small}
  \SetAlCapNameFnt{\small}
}}%
\makeatother

\newcommand{\lstsetCommon}{%
  \lstset{%
    columns=fullflexible,%
    keepspaces=true,%
    escapeinside={<[}{]>},%
    moredelim=**[is][\color{Cerulean}]{<*}{*>},
    moredelim=**[is][\color{Red}]{<^}{^>},
    basicstyle=\ttfamily\footnotesize,%
    showstringspaces=false,%
    aboveskip=0em,%
    belowskip=0em,%
  }%
}

\newcommand{\lstsetMe}{%
  \lstsetCommon{}%
  \lstset{%
  }
}

\lstsetMe{}

\usepackage{soul}

\newif\ifhl{}
\hltrue{}
\hlfalse{}

\ifhl{}
 \newcommand{\chl}[1]{\hl{\textbf{#1}}}
\else
 \newcommand{\chl}[1]{#1}
\fi

\newif\ifdraft{}
\drafttrue{}
\draftfalse{}

\ifdraft{}
  \newcommand{\jhanote}[1]{ {\textcolor{purple} { ***[Shantenu]: #1 }}}
  \newcommand{\mtnote}[1]{ {\textcolor{orange} { ***[Matteo]: #1 }}}
  \newcommand{\ozgurnote}[1]{ {\textcolor{blue} { ***[Ozgur]: #1 }}}
  \newcommand{\tianlenote}[1]{ {\textcolor{yellow} { ***[Tianle]: #1 }}}
  \newcommand{\NOTE}[1]{\phantom{}\begingroup\relax\ifmmode\boldmath\else\bfseries\fi\color{Cerulean}\ignorespaces#1\ignorespaces\endgroup}
  \newcommand{\TODO}[1]{\phantom{}\begingroup\relax\ifmmode\else\sffamily\fi\color{BurntOrange}\ignorespaces#1\ignorespaces\endgroup}
  \newcommand{\FIXME}[1]{\phantom{}\begingroup\relax\ifmmode\boldmath\else\bfseries\sffamily\fi\color{Red}\ignorespaces#1\ignorespaces\endgroup}
  \newcommand{\FIXED}[1]{\phantom{}\begingroup\relax\ifmmode\else\sffamily\fi\color{Green}\ignorespaces#1\ignorespaces\endgroup}
  \newcommand{\DELETE}[1]{\phantom{}\begingroup\relax\ifmmode\else\sffamily\fi\color{Red}\ifmmode\text{\sout{\ensuremath{#1}}}\else\sout{\ignorespaces#1\ignorespaces}\fi\endgroup}
\else
  \newcommand{\jhanote}[1]{}
  \newcommand{\mtnote}[1]{}
  \newcommand{\ozgurnote}[1]{}
  \newcommand{\tianlenote}[1]{}
  \newcommand{\NOTE}[1]{}
  \newcommand{\TODO}[1]{}
  \newcommand{\FIXME}[1]{}
  \newcommand{\FIXED}[1]{}
  \newcommand{\DELETE}[1]{}
\fi

\def\BibTeX{{\rm B\kern-.05em{\sc i\kern-.025em b}\kern-.08em
    T\kern-.1667em\lower.7ex\hbox{E}\kern-.125emX}}

\begin{document}


\title{Workflow Mini-Apps: Portable, Scalable, Tunable \& Faithful Representations of Scientific Workflows}

\newif\ifanon{}
\anontrue{}
\anonfalse{}
\ifanon{}
    \author{
    Anonymous Authors
    }
\else  
    \author{\IEEEauthorblockN{ Ozgur O. Kilic}
    \IEEEauthorblockA{\textit{Brookhaven National Laboratory} \\
    Upton, NY, USA \\
    okilic@bnl.gov, 0000-0003-2129-408X}
    \and
    \IEEEauthorblockN{ Tianle Wang}
    \IEEEauthorblockA{\textit{Brookhaven National Laboratory} \\
    Upton, NY, USA \\
    twang3@bnl.gov, 0000-0001-8293-0671}
    \and
    \IEEEauthorblockN{ Matteo Turilli}
    \IEEEauthorblockA{\textit{Brookhaven National Laboratory} \\
    Upton, NY, USA \\
    \textit{Rutgers-New Brunswick} \\
    Piscataway, NY, USA \\
    mturilli@bnl.gov,0000-0003-0527-1435 }
    \and
    \linebreakand
    \IEEEauthorblockN{ Mikhail Titov}
    \IEEEauthorblockA{\textit{Brookhaven National Laboratory} \\
    Upton, NY, USA  \\
    mtitov@bnl.gov, 0000-0003-2357-7382}
    \and
    \IEEEauthorblockN{ Andre Merzky}
    \IEEEauthorblockA{\textit{RADICAL-Computing Inc.} \\
    Wilmington, DE, USA \\
    andre.merzky@radical-computing.com \\ 
    0000-0002-7228-4327 }
    \and
    \IEEEauthorblockN{Line Pouchard}
    \IEEEauthorblockA{\textit{Brookhaven National Laboratory} \\
    Upton, NY, USA  \\
    pouchard@bnl.gov 0000-0002-2120-6521}
    \and
    \linebreakand
    \IEEEauthorblockN{Shantenu Jha}
    \IEEEauthorblockA{
    \textit{Rutgers-New Brunswick} \\
    Piscataway, NY, USA \\
    \textit{Brookhaven National Laboratory} \\
    Upton, NY, USA  \\
    shantenu@bnl.gov 0000-0002-5040-026X}
    }
\fi

\maketitle


\begin{abstract}
 Workflows are critical for scientific discovery. However, the sophistication,
heterogeneity, and scale of workflows make building, testing, and optimizing
them increasingly challenging. Furthermore, their complexity and
heterogeneity make performance reproducibility hard. In this paper, we
propose workflow mini-apps as a tool to address the challenges in building and
testing workflows while controlling the fidelity of representing 
real-world workflows. Workflow mini-apps are deployed and run on various HPC
systems and architectures without workflow-specific constraints. We offer
insight into their design and implementation, providing an analysis of their
performance and reproducibility. Workflow mini-apps thus advance the science
of workflows by providing simple, portable, and managed (fidelity)
representations of otherwise complex and difficult-to-control real workflows.
\end{abstract}






\section{Introduction}
\label{sec:intro}

Workflows have become an indispensable part of scientific research, particularly with the increased availability 
of high-performance computing (HPC) platforms and the addition of artificial intelligence (AI) and machine learning (ML)~\cite{casalino2020aidriven}. Increased hardware and software heterogeneity~\cite{covidisairborne2021ijhpca}, scale, and sophistication make building, testing, optimizing, and reproducing workflows challenging. 
Additionally, the added dependencies on specific libraries and dedicated hardware complicate deployment to diverse HPC platforms.

Generating workflow mini-apps, i.e., simplified representations of real-world workflows, contributes to addressing challenges from complexity and heterogeneity.  For example, the High Energy Physics Center for Computational Excellence (HEP-CCE) Status and Planning Report~\cite{hepccereport2023indigo,hepccerwebpage} promotes building workflow mini-apps for the ATLAS~\cite{aad2008atlas} and DUNE~\cite{abi2020volume} HEP experiments. According to that plan, workflow mini-apps will provide key performance metrics for future HPC facilities and help the experiment easily leverage new architectures.

Scientists use mini-apps~\cite{fogerty17,sukumar16} to build and test complex applications. They are either simpler single-purpose versions of an entire application, or partial versions of the application's most critical components. When considering workflow applications, single-purpose or partial mini-apps cannot represent the various roles played by tasks within a workflow or the relations among tasks. As such, there is a need to study workflow mini-apps and their specific requirements.

Workflow mini-apps have four main requirements: (1) balancing the trade-off between simplicity and fidelity of representation; (2) simplifying workflow tasks by emulation while maintaining their inter-relationships; (3) enabling portability across different platforms and environments; and (4) enabling performance analysis and reproducibility.

In this paper, we introduce, define, and describe workflow mini-apps. We propose a novel approach to derive mini-apps from workflow applications, and we show the benefits of using those mini-apps for developing, testing, deploying, and optimizing scientific workflows on HPC platforms. We test and validate our proposed approach by deriving mini-apps for two real-world applications: the Inverse Problem~\cite{wang2023parallel} workflow from the ECP ExaLearn project, and an AI-steered Molecular Dynamics(MD) simulation workflow (refered as DeepDriveMD) run using DeepDriveMD~\cite{brace2022coupling} tool from the ECP CANDLE project. 
We chose these two workflows because the Inverse Problem is a representative example of stationary workflow, and DeepDriveMD is an example of adaptive workflow.
We experimentally show that our mini-apps correctly emulate the execution patterns of the original workflow while requiring lower complexity and fewer resources.

We focus our analysis and experiments on four key aspects of workflow mini-apps: scaling, portability, execution models, and performance reproducibility. 
We scale the execution of our mini-apps, measuring their performance and cost in terms of workflow- and task-level makespan, resource utilization and input/output (I/O). \chl{We test our mini-apps portability by measuring their performance characteristics on various HPC platforms with different architectures.} We evaluate synchronous and asynchronous execution models for mini-apps, measuring how they affect performance. Finally, we assess performance reproducibility by measuring run-to-run variation across multiple runs within the same platform and environment~\cite{nicolae2023building,Patki_Thiagarajan_Ayala_Islam_2019}.
Together, our analysis and experiments show that workflow mini-apps have the potential to advance the science and the engineering of workflows by providing simple, portable, and faithful
representations of otherwise complex 
real-world workflows. 

This paper offers five main contributions:
(1) Providing a methodology for building workflow mini-apps.
(2) Deriving workflow mini-apps for two real-world workflow applications (Inverse Problem and DeepDriveMD).
(3) Implementing a 
\chl{Python and C++ library }
with an application programming interface (API) specifically designed 
to allow users to build emulated tasks more easily, based on understanding the bottleneck of the original task.
(4) Providing extensive experimental performance comparisons of the mini-apps and their workflows on Polaris
, Summit and Frontier
, three platforms of the Department of Energy Leadership Class Computing Facilities. 
(5) Showing how workflow mini-apps can be utilized in a more general context for portability and reproducibility. 

\begin{figure}[h]
\centering
   \includegraphics[width=0.45\textwidth]{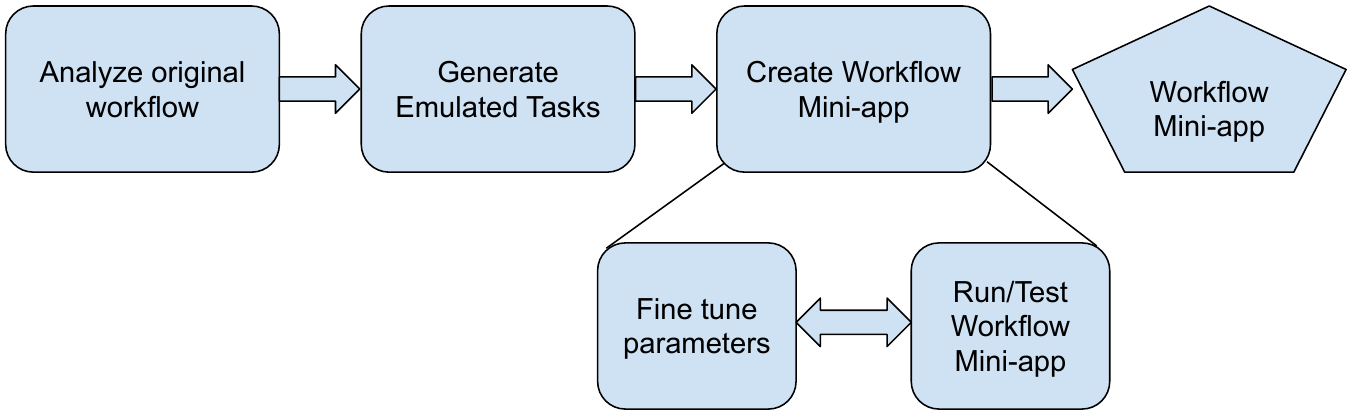}
\caption{Design process of a workflow mini-app}
\label{fig:design}
\end{figure}

\section{Related Works}
\label{ser:related}

Exascale Computing Project (ECP) proxy applications~\cite{ecpProxy,matsuoka2022preparing} have been developed and used to show important features of large applications in the HPC area. There are many proxy apps, mini-apps, and benchmarks that have been built as part of ECP, such as ECP CANDLE Benchmarks~\cite{wu2019performance}, ExaMiniMD~\cite{thompson2018ecp}, and miniVite~\cite{ghosh2018minivite}. 
However, they all focus on a single application. In this paper, we focus on workflows instead, where workflows are a combination of various interrelated tasks/applications.

Benchmark suites are used for developing or testing scientific applications or HPC systems. NERSC-10 Benchmark Suite~\cite{nersc} plays a critical role in evaluating the systems with micro-benchmarks and workflow component benchmarks. Their benchmarks represent functions/tasks of the workflows that are helpful for testing different systems for critical tasks and their performance.  However, benchmarks are not sufficient for our purposes as they are more problem-specific while mini-apps have been used for both testing the performance of the workflow application and for portability.

Mini-apps also played a critical role in scientific computing in different areas. A molecular docking mini-app is used to validate the tunable approximation approach proposed in Ref.~\cite{gadioli2021tunable}. Neural Mini-Apps~\cite{vineyard2022neural} are used to get a broader understanding of Neuromorphic computing (NMC). The arch project~\cite{martineau2017arch} provides a suite of mini-apps for evaluating physics algorithms. Neuromapp~\cite{ewart2017neuromapp} is a framework that focuses on single-functionality mini-apps to improve neural simulators. In data science~\cite{sukumar16}, mini-apps are used to identify data-parallel kernels and test different programming models. These references argue that mini-apps should enable the exploration of the performance of the original application while providing simplicity and portability, as we show in this paper.

Our workflow mini-apps have the same common characteristics as the mini-apps in the literature. However, workflow mini-apps differ from them as they focus on workflows as a whole instead of focusing on single-functionality or application. For example, workflow mini-apps can provide a more realistic picture of resource competition and communication between its components (tasks), and the effect comes from task dependency where regular mini-apps cannot.
Workflow mini-apps can provide needed performance insight of workflow where single-purpose mini-apps could not.

Performance reproducibility of the workflows is challenging~\cite{ramesh2022ghost, nicolae2023building,pouchard2019computational} due to both the nature of workflows (complex data- and compute-intensive, AI/ML) and the platform on which they are running (system reliability, unsuccessful execution, performance fluctuation of the file system and the network).
Performance reproducibility of full applications is costly and non-trivial to achieve, a challenge addressed by workflow mini-apps that offer performance fidelity while providing simplicity and reduced costs.



\section{Design Workflow Mini-Apps}
\label{sec:design}
\jhanote{Design of what? We begin with "recipe" ... }\mtnote{I agree and suggested an iteration}\ozgurnote{I agree}

We provide a methodology to design workflow mini-apps for any given workflow. 
We discussed in \S~\ref{sec:intro} that scientific workflows are hard to develop, deploy, or run, and we ask the following questions:
(1) What are the essential characteristics of a given workflow that should be implemented into its mini-app?
(2) What would we need to enable performance reproducibility within the workflow mini-apps? 
(3) What are the target use cases?

In terms of the workflow characteristics, we focus on performance. Our workflow mini-apps can be used to understand the performance of the original workflow; however, they will not provide any scientifically relevant results. While there are many performance metrics that are useful, we focus on three main metrics: makespan, resource utilization, and I/O utilization at both workflow- and task-level. A combination of these three metrics can be used to understand performance and find possible bottlenecks~\cite{pathway2013,benedict2013performance,luttgau2018toward,mattoso2015dynamic}. It is known that memory (accessing data) is also one of the most significant bottlenecks in HPC applications~\cite{kilic2022memgaze,kilic2020rapid}. However, we choose not to focus on memory utilization as one of the metrics for two main reasons. First, memory utilization depends on applications' access patterns of the memory (algorithm dependent), which makes it application (i.e., task) specific.
Second, memory utilization (memory bottleneck) is also architecture (CPU/Accelerator) specific, which makes it not portable. Instead, we focused on makespan, which includes any bottlenecks (including memory) and can be applied to the entire workflow.

While this work focuses on the above three performance metrics, users are free to add any additional metric into their own design, as long as the workflow middleware is compatible with the profiler that collects it. For example, as a future work, 
we will integrate metrics of
performance anomaly, 
collected via Chimbuko\cite{kelly2020chimbuko}, a performance anomaly detection system. 

\begin{figure}[h]
\centering
\includegraphics[width=0.3\textwidth]{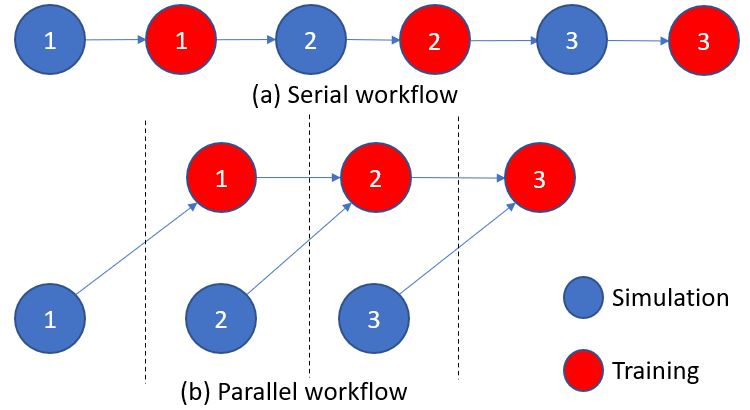}
  \caption{The two workflows used in the Inverse Problem. Here, we take the \#phase to be three as an example. The number inside each task represents the phase-id the task belongs to, and the dashed line is the boundary of different stages, while tasks in different stages can not run in parallel. The arrow represents task dependency: A Task will not start until all tasks that point to it are finished.~\cite{wang2023parallel}}
\label{fig:rct-workflow-exalearn}
\end{figure}

We also consider performance reproducibility as part of our design.  Performance reproducibility is the minimal run-to-run variation across multiple runs of the same application, using a consistent set of configurations on the same system~\cite{Patki_Thiagarajan_Ayala_Islam_2019}.
As reproducing executions of applications on the same architecture still shows much performance variability~\cite{nicolae2023building}, workflow mini-apps can be used to analyze variations and point a direction toward deeper root-cause analysis.  The increased complexity of workflows is exemplified in workflow patterns, execution on heterogeneous architectures, hybrid orchestration of numerical and data-intensive simulations, and execution that is often non-deterministic (e.g., ML-driven MD simulations\cite{brace2022coupling}). \jhanote{Can you give examples of the complexity of workflows?}\ozgurnote{done} This complexity introduces new challenges for performance reproducibility:  workflow management systems do not always have mechanisms to collect workflow behavior, workflow performance overhead, and application performance.  A lack of metadata and provenance related to access to the original input data, execution scripts, and runtime environment~\cite{nicolae2023building} are barriers to reproducibility.\jhanote{How are these barriers to reproducibility?}\ozgurnote{Line addressed this issue.}

With the target use cases explained in \S~\ref{sec:intro}, in our design, we focus on the workflow as a whole instead of individual tasks within the workflow. Fig.~\ref{fig:design} shows the design of our workflow mini-apps, which has a three-step approach. First, we must analyze and understand the original workflow and collect performance data (resource utilization, I/O, and makespan). 
Second, we generate emulated tasks using our API library~\cite{mini-apps}. The library provides a large set of simple but representative kernels and corresponding parameters (Table.~\ref{tab:wfminiapi}), allowing users to assemble them to build various tasks with different bottlenecks. 
Finally, we built the workflow using emulated tasks and a workflow middleware, RADICAL Cybertools (RCT)\cite{balasubramanian2019radical,entk2016bala}. We chose RCT simply because both original workflows were using RCT. In reality, any workflow middleware (e.g., PanDA~\cite{maeno2011overview}, Colmena~\cite{ward2021colmena}) can be used for developing the workflow mini-app, and we highly recommend developers choose the workflow middleware that they choose to implement the original workflow, since the implementation of emulated tasks based on the API library is completely independent of the workflow middleware (except for the inter-task communication).
To ensure the performance fidelity of the original workflow, we ran the workflow mini-app and fine-tuned it by adjusting the parameters we provided. 
In the next section, we will describe each step in detail.

\begin{figure}[h]
\centering
   \includegraphics[width=0.3\textwidth]{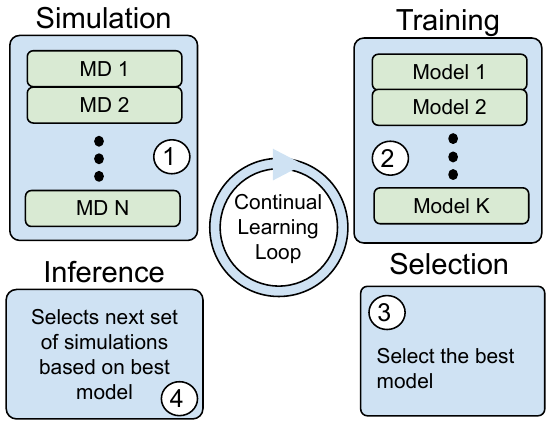}
\caption{This figure shows the DeepDriveMD overview for synchronous execution model~\cite{brace2022coupling}.}
\label{fig:ddmd}
\end{figure}

\section{Implementation}
\label{sec:implementation}

As outlined in \S~\ref{sec:design}, our design includes three stages shown in Fig~\ref{fig:design}.
This section explains each stage in detail and the implementation of both Inverse Problem and DeepDriveMD workflow mini-apps.

The Inverse Problem workflow, which is part of the ExaLearn project~\cite{wang2023parallel,alexander2021co}, is designed to assess Neutron diffraction outcomes, specifically the Bragg profile, to predict the symmetry class of a material along with evaluating its unit cell parameters. To summarize, the process begins by generating simulation data (the simulation task). This data is then used to train a deep learning model that predicts both the symmetry class and unit cell parameters (the training task). This iterative procedure is repeated multiple times(phases) to refine and enhance the model's performance with each phase. Fig.\ref{fig:rct-workflow-exalearn} illustrates both the sequential (Fig.\ref{fig:rct-workflow-exalearn}a) and parallel (Fig.~\ref{fig:rct-workflow-exalearn}b) workflows, highlighting their respective dependencies.

DeepDriveMD~\cite{brace2022coupling,lee2019deepdrivemd}, on the other hand, is a deep-learning (DL) oriented molecular dynamics (MD) simulation workflow. MD simulations have broad applications, such as probing intricate biophysical phenomena like protein folding, protein-ligand, or small molecule docking and fostering crucial advancements in areas like drug design. Fig.~\ref{fig:ddmd} showcases its four primary components: (1) MD simulation, (2) ML training for the generation of ML models, (3) Model Selection, and (4) Inference, which aids in determining the subsequent set of simulation starting points.

\subsection{Analyzing Original Workflow}
\label{sec:impl:analysis}

Analyzing the foundational structure and intricacies of the original workflow is crucial when developing workflow mini-apps. This involves understanding the workflow's various components and their interaction. Our approach to analyzing the workflow is tri-fold: (1) Identifying and categorizing every task type in the workflow, (2) understanding the execution model, focusing predominantly on the dependencies and intercommunication among tasks, and (3) extracting performance metrics for every task.

Within the Inverse Problem and DeepDriveMD workflows, we have two and four distinct task categories, respectively. Tasks within the same category exhibit similarities: they are launched from identical binaries or scripts, implying the same algorithms and consistent operational behavior. Crucially, tasks of the same category often face identical performance bottlenecks. Furthermore, tasks within each category are self-contained, serving a singular purpose, as explained at the beginning of this section. Consequently, the most optimized strategy is constructing two and four emulated tasks for the Inverse Problem and DeepDriveMD workflows, respectively.

The dependencies and communications among tasks will help us understand the execution models of the workflow and possibly allow us to improve it. Here, we take the Inverse Problem as an example. As shown in Fig.~\ref{fig:rct-workflow-exalearn}, in the serial workflow, a training task in phase\_$i$ (train\_$i$) depends on the simulation task in phase\_$i$ (sim\_$i$), and sim\_$i$ depends on train\_($i-1$). Because of that, all tasks are running in serial in this workflow. In the parallel workflow, we make extra optimization so that sim\_$i$ does not depend on train\_($i-1$) anymore, suggesting that sim\_$i$ and train\_($i-1$) can run in parallel, but train\_$i$ still depend on train\_($i-1$), which means all \textit{training} tasks have to run in serial.

Finally, collecting performance metrics is essential to understand the execution characteristics of each task. As mentioned in \S~\ref{sec:design}, we collect makespan, resource utilization, and I/O size. These metrics have proven helpful for generating performance characteristics of each task within the workflow.
In this work, we use profiling tools, including RADICAL Analytics~\cite{radical_analytics_url} to collect resource utilization and makespan, and Darshan~\cite{carns200924} to collect I/O utilization. 
For ease of use, we also provide the user with the corresponding analyzing scripts.

\subsection{Generating Emulated Tasks}
\label{sec:imp:mini-tasks}

As explained in \S.~\ref{sec:intro} and \S~\ref{sec:design}, tasks within the workflow mini-apps should be black boxes that are simple and emulate original workflow tasks without dependencies on libraries that require a specific operating system (OS) or  Central Processing Unit(CPU) (e.g., PowerPC, ARM, and x86)/Graphical Processing Unit(GPU) (e.g., NVIDIA and AMD) architectures.
Also, 
emulated task design must be tunable since different tasks of the same category could have different configurations, giving different performance behaviors.

To address the above issues, we designed an open-source intermediate library, wfMiniAPI\cite{mini-apps}. This library provides users with a set of APIs that target different functionalities, including various computation kernels on CPU and GPU with different bottlenecks, MPI communications, data movements, and I/O operations. Since most tasks are written in Python and C++, we provide both Python and C++ interfaces. wfMiniAPI makes it easier for users to build emulated tasks.

\begin{table}[h]
\centering
\caption{\chl{wfMiniAPI: Some of the APIs used in this work with their tunable parameters, and the emulated tasks called them. API with \textasciicircum symbol means it does not have Python support (only C++) }
}
\begin{tabular}{|p{2.9cm}|p{2.4cm}|p{2.3cm}|}
\hline
\textbf{API} & \textbf{Parameters} & \textbf{Emulated task used}\\
\hline\hline
readNonMPI & data\_size & All tasks\\
\hline
writeNonMPI & data\_size & All tasks\\
\hline
readWithMPI & data\_size & \\
\hline
writeWithMPI & data\_size & \\
\hline\hline
MPIallReduce & device, data\_size & Training\\
\hline
MPIallGather & device, data\_size & \\
\hline
MPIallReduceAsync\textasciicircum & device, data\_size & \\
\hline\hline
matMulGeneral & device, dim\_list & Training\\
\hline
matMulSimple2D & device, dim & Simulation\\
\hline
fft & device, data\_size, type\_in, transform\_dim & \\
\hline
RNG & device, data\_size, distribution & \\
\hline
axpy ($y=a\cdot x + y$) & device, data\_size & MD simulation\\
\hline
scatterAdd ($y[idx[i]] += x[i]$) & device, x\_size, y\_size & \\
\hline
reduction & device, data\_size & \\
\hline
inplaceCompute ($y[i] = f(y[i])$) & device, functor, data\_size & \\
\hline\hline
dataCopyD2H & data\_size & MD simulation, Training, Agent\\
\hline
dataCopyH2D & data\_size & MD simulation, Training, Agent\\
\hline
dataCopyD2HAsync\textasciicircum & data\_size & \\
\hline
dataCopyH2DAsync\textasciicircum & data\_size & \\
\hline
\end{tabular}
\label{tab:wfminiapi}
\end{table}

As for the implementation of wfMiniAPI, the Python interface is implemented with packages of Numpy, Cupy, mpi4py, and h5py, which are currently provided by or can be easily installed on almost all common HPC clusters. The C++ interface is implemented with MPI and OpenMP offloading to target CPU and NVIDIA/AMD GPU. Here, we chose OpenMP offloading instead of other programming models like SYCL or Kokkos, mainly because it is usually the easiest to deploy and widely available in most LCF/HPC machines. This makes emulated tasks that are built with this library portable. 

In addition, each API in wfMiniAPI accepts parameters that allow them to be tunable. Users can choose appropriate APIs according to the original task, assemble them to build emulated tasks, and adjust parameters for different task configurations. For example, the simulation tasks in the Inverse Problem are MPI applications with one thread per process and negligible inter-process communication, and each process mainly does regressions (with a bottleneck of matrix multiplication) and disk I/O. We can use the code snippet in Listing~\ref{code:sim} as an emulated task for the simulation task. Here, num\_data, read\_size, num\_mult, mat\_size, and write\_size are all tunable parameters. Table.~\ref{tab:wfminiapi} \chl{provides kernels within wfMiniAPI with most important parameters for each kernel and their use-cases. Since we design wfMiniAPI to be extendable for a current list of the kernels we refer the reader to our GitHub} repo\cite{mini-apps}.

\vspace{5pt}
\begin{lstlisting}[style=mystyle, label=code:sim, caption=Main component of the emulated simulation task in the Inverse Problem workflow mini-app.][tbp!]
for i in range(num_data):
    wfMiniAPI.readNonMPI(read_size)
    for j in range(num_mult):
        wfMiniAPI.matMulSimple2D(mat_size)
    wfMiniAPI.writeNonMPI(write_size)    
\end{lstlisting}
\vspace{5pt}

Since wfMiniAPI is open source, users can contribute by adding their customized API. With the wfMiniAPI library, users can easily build emulated tasks without needing to write most of the code while getting performance metrics comparable to the original task. It also allows users to have decent code reuse: One user's contribution to the API list can be reused by other users in developing other emulated tasks, and code for building one workflow mini-app might be used for building another workflow mini-app that shares similar tasks.

One question people often ask is: Why not use a simple sleep function with a tunable sleep time as an emulated task? This is because our workflow mini-app is not simply going to analyze if the execution model of the original workflow can run successfully with the workflow middleware the developer chooses, but we also care about performance analysis of the original workflow, especially under different setups, including different problems inputs, and different platform/hardware architectures. However, with different setups, the running time of each sub-task could vary a lot, and it is usually hard to theoretically predict quantitatively how their running time would vary and use that number as the parameter of the sleep function. Because of that, using sleep functions as emulated tasks is no longer an option, and we need to create emulated tasks that are composed of a small number of kernels that represent the bottleneck of the original tasks, and it is the only way to estimate and predict the running time of the workflow under different setups.

\vspace{-2pt}
\subsection{Creating Workflow Mini-apps}
\label{sec:imp:wfminiapps}
\mtnote{I edited the subsection more than I had originally planned. That saved a fair amount of space but pls check that I did not altered the intended meaning.}
\ozgurnote{Thank you this looks perfect}

As mentioned in \S~\ref{sec:design}, we use pilot-based execution middleware (RCT), 
separating the workflow execution model and the details of task submission. This simplifies the development of the mini-apps since many workflows are similar, and users can use templates, customizing them to generate the workflow with the desired execution model without implementing a new workflow mini-app from the beginning.

After building the skeleton of the workflow, the next step is to fine-tune the parameters 
introduced when  
creating emulated tasks so that the workflow mini-app can generate performance metrics comparable to the original workflow in \textit{any} configuration. Here, comparable means that the original workflow and the workflow mini-app should have similar resource utilization, and the ratio between their makespan and I/O size should be fixed.

\mtnote{I rewrote the following, but then decided that it should be commented out. Pls check that you agree.}
\ozgurnote{I agree we can remove the paragraph if we need space}

We perform fine-tuning by choosing 
a base configuration with some specific problem input and architecture, and running the original workflow once to collect the relevant performance metrics. We then tune the parameters of the workflow mini-app so that its performance metrics are comparable to those of the original workflow base configuration but proportionally several times smaller. For example, the tuning include kernels parameters in the API library and the number of kernel calls in emulated tasks.
Usually, most of the parameters of an emulated task can be fine-tuned based on some domain knowledge. For example, the number of epochs, the number of matrix multiplications, the size of matrix multiplication or the size of data I/O. 
\chl{Completing the fine-tuning requires running the workflow mini-app 3-5 times for the two examples in this work.}

After fine-tuning, we find the relation of the parameters between the original workflow and the workflow mini-app, and generate a mapping between them. In that way, for any other configurations of the original workflow, we do \textbf{not need} to repeat the fine-tuning process; we only adjust the parameters needed for the new configuration, using the mapping generated during the fine-tuning process. 
We will show the correctness of this process in \S.~\ref{sec:eval_validation}.

\section{Evaluation}
\label{sec:evaluation}

\begin{table*}[h]
\centering
\setlength\tabcolsep{4.7pt}
\caption{Validation experiments setup. Inverse Problem (IP) mini-app (m-app): 3 configurations V1--3; varying input size (Data column) and number of epochs. DeepDriveMD (DDMD) m-app: 2 configurations V1--2; varying the MD simulation length (Steps column), and the number of epochs and phases. Mini-apps configurations represent those used for scientific measurements (e.g., DDMD data 1FME~\cite{sarisky2001betabetaalpha}), enabling fidelity validation for both workflows at different problem sizes, i.e., configurations V1--3.
\jhanote{I do not see mention of V1, V2 ...?} \mtnote{Uff, that required dealing with the beautiful latex, such an amazing tool\ldots On another note, why are we calling a configuration `V' instead of `C'?}\ozgurnote{I think we started with version and V then stuck with it we can change it to C}
\jhanote{boldface the first row? Try to reduce the caption size if possible.}\mtnote{Done. Pls check and if you agree/like, change the style of the other two tables. Note the abbreviations IP and DDMD are necessary because the table was too large. I made them consistent across the tables but need to be explicitly states in each caption.}\ozgurnote{Thank you, I think it is great}}
\begin{small}
\begin{tabular}{|lll|c|c|c|ccc|ccc|ccc|cc|cc|}
\hline
\multicolumn{3}{|l|}{\multirow{2}{*}{\textbf{Execution Model}}} & 
\multirow{2}{*}{\textbf{\#Node}}          & 
\multirow{2}{*}{\textbf{\#CPU}}           & 
\multirow{2}{*}{\textbf{\#GPU}}           & 
\multicolumn{3}{|c|}{\textbf{\#Rank}}     &
\multicolumn{3}{|c|}{\textbf{\#Epoch}}    &
\multicolumn{3}{|c|}{\textbf{Data}}       &
\multicolumn{2}{|c|}{\textbf{\#Phase}}    &
\multicolumn{2}{|c|}{\textbf{\#Step}}    \\
\cline{7-19}
& &           &
              &
              &
              &
\textbf{sim}  &  
\textbf{ml}   &  
\textbf{rest} &  
\textbf{V1}   &  
\textbf{V2}   &  
\textbf{V3}   &  
\textbf{V1}   &  
\textbf{V2}   &  
\textbf{V3}   &  
\textbf{V1}   &  
\textbf{V2}   &  
\textbf{V1}   &  
\textbf{V2}   \\ 
\hline
IP & Serial & CPU &  
4                 &  
128               &  
0                 &  
128 & 4  & -      &  
100 & 50 & 50     &  
X   & X  & 2X     &  
3   & -           &  
-   & -           \\ 
\hline
m-app & Serial & CPU & 
4                 & 
128               &
0                 &
128 & 4  & -      &
50  & 25 & 25     &
X   & X  & 2X     &
3 & -             &
- & -             \\
\hline
IP & Serial & CPU+GPU & 
4                     & 
128                   &
16                    &
128  & 16  & -        &
1600 & 800 & 800      &
X    & X   & 2X       &
3 & -                 &
- & -                 \\
\hline
m-app & Serial & CPU+GPU & 
4 &
128 &
16 &
128 & 16 & - &
200 & 100 & 100 &
X & X & 2X &
3 & - &
- & - \\
\hline
IP & Parallel & CPU &
8 & 
256 &
0 &
128 & 4 & - &
100 & 50 & 50 &
X & X & 2X &
3 & - &
- & - \\
\hline
m-app & Parallel & CPU &
8 &
256 &
0 &
128 & 4 & - &
50 & 25 & 25 &
X & X & 2X &
3 & - &
- & - \\
\hline
\multicolumn{3}{|l|}{DDMD}  &
3  &
96 &
12 & 
12 & 1 & 1  &
100  & 150 & - &
1FME & 1FME   & - &
2 & 3 & 
4k & 5k\\
\hline
\multicolumn{3}{|l|}{m-app}  &
3  &
96 &
12 &  
12 & 1 & 1  &
100 & 150 & - &
- & - & - &
2 & 3 & 
4k & 5k \\
\hline
\end{tabular}
\label{tab:exp-setup}
\vspace{-10pt}
\end{small}
\end{table*}

\mtnote{I edited the opening paragraph making it more concise (we need to cut 1 page!) but also making explicit that the scope of our experiments is `limited' to the two exemplar mini-apps, not every mini-app. I think the results of the experiments need to be then generalized in their discussion/analysis. Pls check and revert if you disagree.}\ozgurnote{I think this looks good}

In this section, we present the evaluation of workflow mini-apps by performing five sets of experiments with the DeepDriveMD and Inverse Problem exemplar workflow mini-apps implemented in \S~\ref{sec:implementation}. For those two workflow mini-apps, our experiments: (1) validate their fidelity; (2) prove that they can enable performance reproducibility; (3) show their portability across platforms with different architectures; (4) illustrate how they can facilitate testing scalability; and (5)  show that they can serve as a testbed for workflow optimization.

\begin{table*}[hb]
\centering
\setlength\tabcolsep{4pt}
\caption{Results for our validation experiments with different execution models and experiment configurations in Table~\ref{tab:exp-setup}. Multiple numbers within each cell represent different configurations (three for the Inverse problem, two for the DeepDriveMD). Execution time is the total execution time of the workflows. Similar to execution time, CPU and GPU utilization is provided for the entire runtime of the workflows.}
\begin{small}
\begin{tabular}{|lll|ccc|ccc|ccc|ccc|ccc|}
\hline
\multicolumn{3}{|l|}{\multirow{2}{*}{\textbf{Experiment}}} &
\multicolumn{3}{|c|}{\textbf{Makespan (s)}} &
\multicolumn{3}{|c|}{\textbf{CPU (\%)}} &
\multicolumn{3}{|c|}{\textbf{GPU (\%)}} &
\multicolumn{3}{|c|}{\textbf{Read I/O (GB)}} &
\multicolumn{3}{|c|}{\textbf{Write I/O (GB)}} \\
\cline{4-18}
& & &
\textbf{V1}   &  
\textbf{V2}   &  
\textbf{V3}   &  
\textbf{V1}   &  
\textbf{V2}   &  
\textbf{V3}   &  
\textbf{V1}   &  
\textbf{V2}   &  
\textbf{V3}   &  
\textbf{V1}   &  
\textbf{V2}   &  
\textbf{V3}   &  
\textbf{V1}   &  
\textbf{V2}   &  
\textbf{V3}   \\  
\hline
IP     & Serial & CPU    & 
1840.3 & 1336.3 & 2527.6 &
100    & 100    & 100    &
0      & 0      & 0      &
560.5  & 560.8  & 1115.0 &
126.2  & 126.2  & 252.2  \\
\hline
m-app & Serial & CPU &
428.3 & 327.0 & 572.8  &
100 & 100 & 100 &
0 & 0 & 0 &
128.2 & 128.1 & 257.0 &
29.0 & 29.0 & 59.6 \\
\hline
IP     & Serial & CPU+GPU &
2148.9 & 1316.2 & 2098.8  &
100    & 100    & 100     &
62     & 50     & 40      &
679.2  & 678.8  & 1348.6  &
127.0  & 127.1  & 252.2.  \\
\hline
m-app & Serial & CPU+GPU &
462.2 & 368.3 & 571.5 &
100 & 100 & 100 &
60 & 50 & 44 &
156.8 & 156.8 & 313.2 &
28.9 & 28.9 & 59.6\\
\hline
IP & Parallel & CPU &
1503.7 & 869.7 & 1748.0 &
63 & 72 & 73 &
0 & 0 & 0 &
560.5 & 560.7 & 1115.0 &
126.0 & 125.9 & 252.3\\
\hline
m-app & Parallel & CPU &
326.0 & 222.9 & 395.1 &
64 & 70 & 69 &
0 & 0 & 0 &
128.9 & 128.7 & 257.0 &
29.0 & 29.0 & 59.5\\
\hline
\multicolumn{3}{|l|}{DDMD}  & 
1479.0 & 3062.3 & -              &
39     & 63     & -              &
39     & 62     & -              &
368.3  & 594.7  & -              &
170.6  & 306.6  & -              \\
\hline
\multicolumn{3}{|l|}{m-app} &
780.1 & 1442.8  & -  &
43 & 67  & - &
43 & 62  & - &
182.4 & 294.8  & - &
84.1 & 148.3  & - \\
\hline
\end{tabular}
\label{tab:exp-results}
\end{small}
\end{table*}

\begin{figure*}[hb]
\centering
    \subfloat[RU DeepDriveMD original workflow V2]{\includegraphics[width=0.32\textwidth]{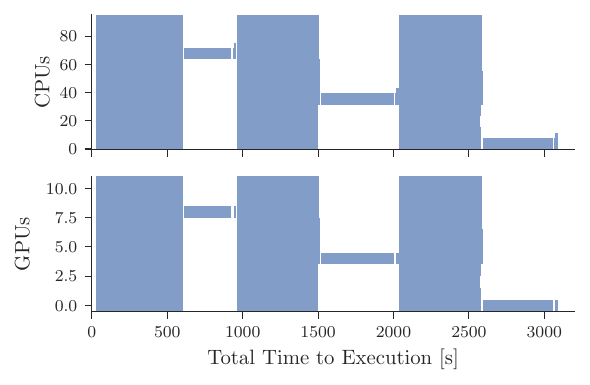}\label{fig:orig_ru_ddmd}}\hfill
    \subfloat[RU Inverse Problem original workflow V2]{\includegraphics[width=0.33\textwidth]{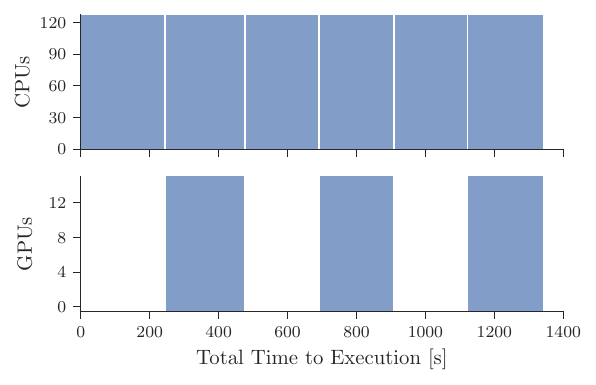}\label{fig:orig_ru_v2}}\hfill
    \subfloat[RU Inverse Problem original workflow V3]{\includegraphics[width=0.33\textwidth]{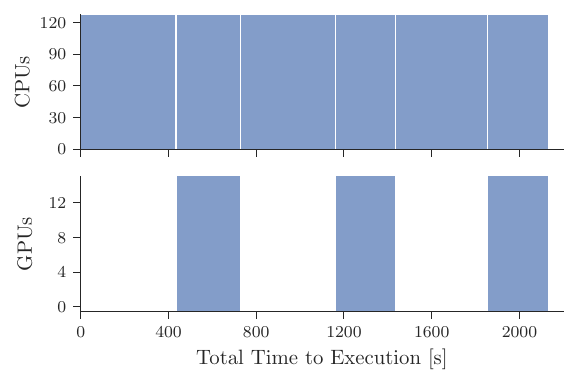}\label{fig:orig_ru_v3}}\hfill
    \\
    \subfloat[RU DeepDriveMD workflow mini-app V2]{\includegraphics[width=0.32\textwidth]{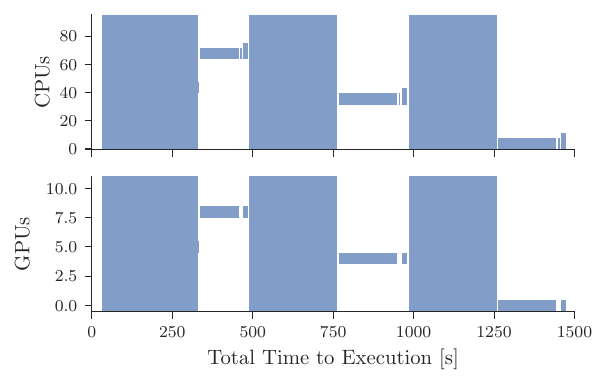}\label{fig:wf_mini-app_ru_ddmd}}\hfill
    \subfloat[RU Inverse Problem workflow mini-app  V2]{\includegraphics[width=0.33\textwidth]{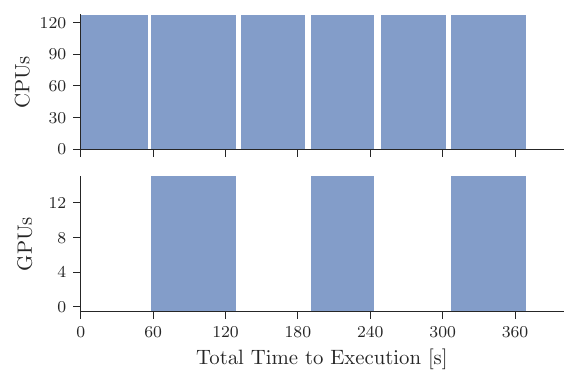}\label{fig:wf_mini-app_ru_v2}}\hfill
    \subfloat[RU Inverse Problem workflow mini-app V3]{\includegraphics[width=0.33\textwidth]{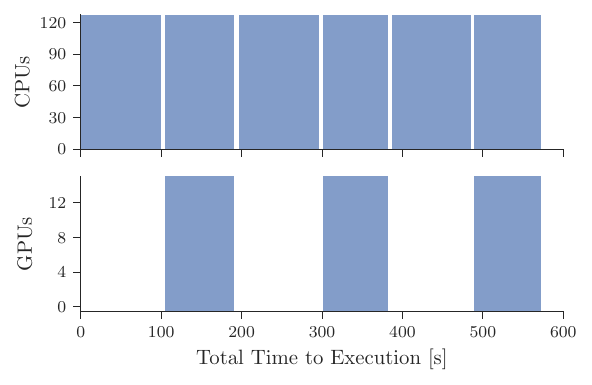}\label{fig:wf_mini-app_ru_v3}}\hfill
\caption{The CPU and GPU resource utilization (RU) as a function of execution time of the original workflow and the workflow mini-app for DeepDriveMD (column 1) and Inverse Problem with an execution model of CPU+GPU serial (column 2 for configuration V2 and column 3 for configuration V3, respectively).  The top three figures show the RU of original workflows and the bottom three figures show the RU of workflow mini-apps. We compare the original workflow with the workflow mini-app for each experiment configuration to validate the fidelity of the workflow mini-apps. \chl{Note that CPUs and GPUs in y-axis refer to CPU and GPU id respectively.}
}
\label{fig:originalWF-ru}
\end{figure*}

\begin{figure*}[hb]
\centering
    \subfloat[I/O DeepDriveMD original workflow V2]{\includegraphics[width=0.32\textwidth]{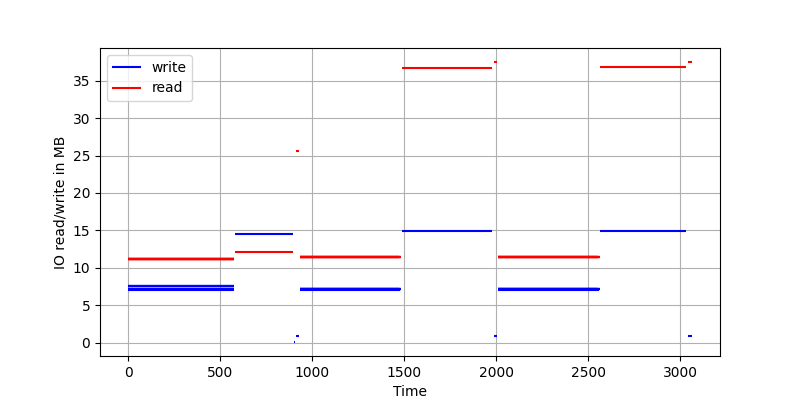}\label{fig:orig_io_ddmd}}
    \subfloat[I/O Inverse Problem original workflow V2]{\includegraphics[width=0.33\textwidth]{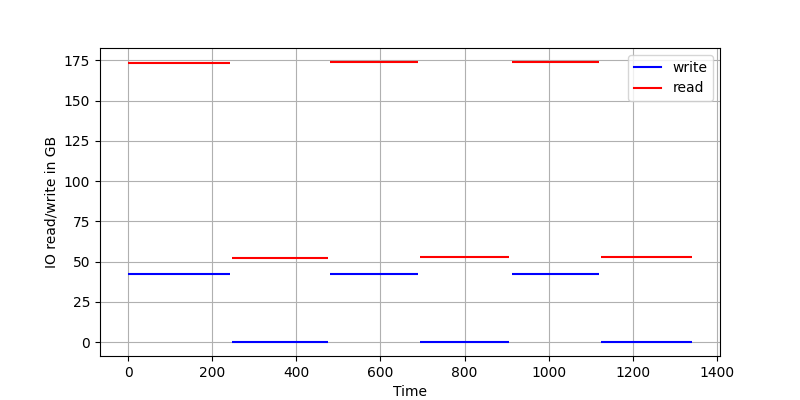}\label{fig:orig_io_v2}}
    \subfloat[I/O Inverse Problem original workflow V3]{\includegraphics[width=0.33\textwidth]{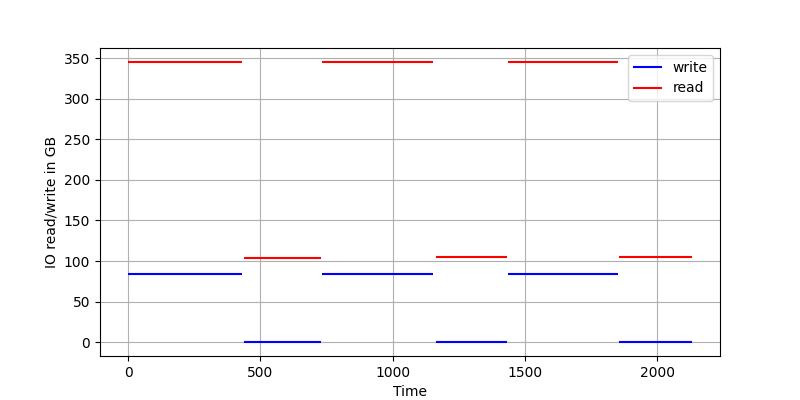}\label{fig:orig_io_v3}}
    \\
    \subfloat[I/O DeepDriveMD workflow mini-app V2]{\includegraphics[width=0.32\textwidth]{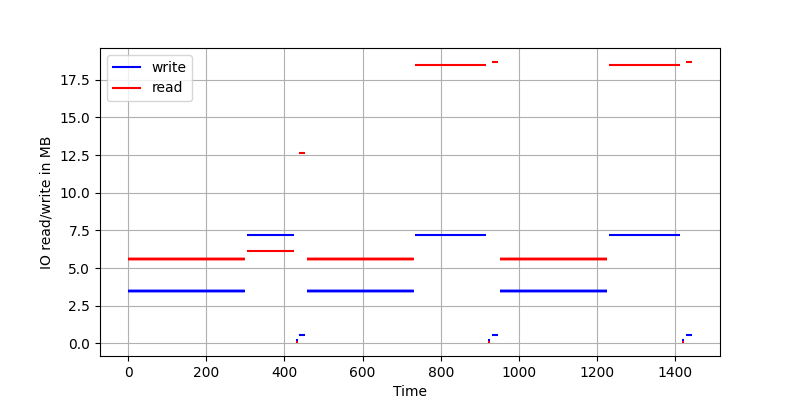}\label{fig:wf_mini-app_io_ddmd}}
    \subfloat[I/O Inverse Problem workflow mini-app V2]{\includegraphics[width=0.33\textwidth]{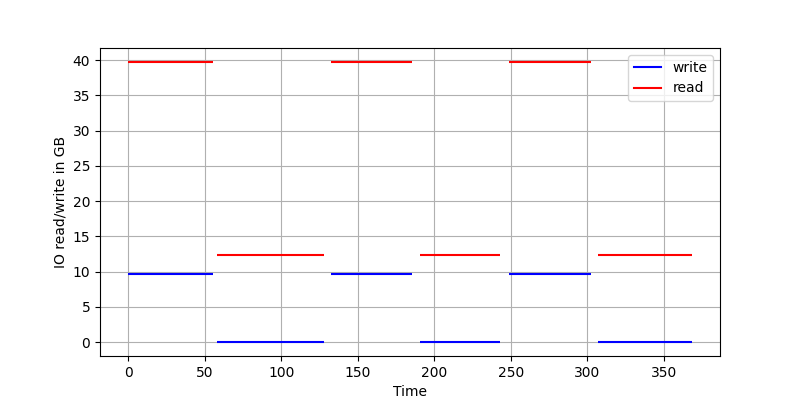}\label{fig:wf_mini-app_io_v2}}
    \subfloat[I/O Inverse Problem workflow mini-app V3]{\includegraphics[width=0.33\textwidth]{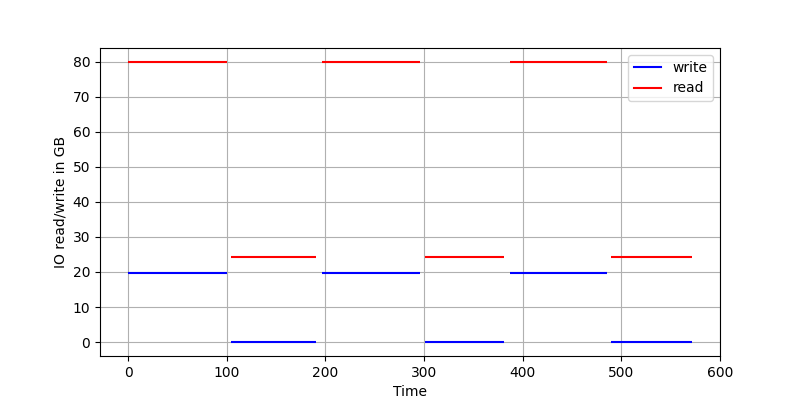}\label{fig:wf_mini-app_io_v3}}
\caption{The I/O performance of the original workflow and the workflow mini-app for DeepDriveMD (column 1) and Inverse Problem with an execution model of CPU+GPU serial (column 2 for configuration V2 and column 3 for configuration V3, respectively). Each plot shows the total read/write size of each task in that workflow during their runtime: \chl{Each line segment has its left and right end representing the start and end time of a single emulated task, and its y-value being the size of I/O of that emulated task.} The top three figures show the original workflow, and the bottom three figures show the workflow mini-app. We compare the original workflow with the workflow mini-app for each experiment configuration to validate the fidelity of the workflow mini-apps.
}
\label{fig:originalWF-io}
\end{figure*}

\subsection{Validation}
\label{sec:eval_validation}

We validate our workflow mini-apps by measuring and comparing the three metrics explained in \S~\ref{sec:design} (makespan, resource (CPU/GPU) utilization, and read and write I/O) for the actual workflows and for their mini-apps. The validity of workflow mini-apps depends on having a controlled fixed ratio between the workflow and its mini-app for each performance metric for \textit{any} configuration. This ratio should be determined based on cost-saving requirements. We used ALCF Polaris
for all the validation experiments because the original Inverse Problem workflow could only run on that platform.

Table~\ref{tab:exp-setup} shows the experimental setup for both workflows. We validate the Inverse Problem (IP) with the three execution models shown in Fig.~\ref{fig:rct-workflow-exalearn}: serial CPU, serial CPU+GPU where the training task runs on GPU, and parallel CPU.
For each execution model, 
we ran three experiment configurations (V1--3), toggling the number of epochs and the input data size.
\chl{We use the combination of serial CPU and configuration V1 as the base configuration for tuning the parameters in the workflow mini-app and adjust the parameters for different configurations V, based on its mapping.} 

For DeepDriveMD, we run two experiment configurations (V1--2) that differ in the number of epochs, length of MD simulation, and number of phases. We use the first configuration for tuning parameters in the workflow mini-app and adjust parameters for the second configuration based on its mapping. There are, respectively, 13 and 21 tuneable parameters for the Inverse Problem and DeepDriveMD workflow mini-apps. 
We enable reproduction by publishing the full data set in a dedicated GitHub~\cite{mini-apps} repository.

Table~\ref{tab:exp-results} summarizes our experiment results. Note that, as a design choice, we used a fixed ratio between workflow and mini-app to reduce both the makespan and amount of read and written data by the mini-app. That reduced the cost of running the workflow mini-apps, i.e., one of the main reasons to use a mini-app. When comparing the results in Table~\ref{tab:exp-results}, we focus on the running time and I/O ratio for each run of the workflow mini-apps and the original workflows. We define the ratio in Eq.~\ref{eq:ratio}:
\begin{equation}
R^{t}_i = \frac{Time^{miniapp}_{conf_i}}{Time^{workflow}_{conf_i}}
    \quad\mathrm{and}\quad 
R^{r/w}_i = \frac{Read/Write^{miniapp}_{Conf_i}}{Read/Write^{workflow}_{conf_i}}
\label{eq:ratio}
\end{equation}

Take the ``serial CPU'' model from the Inverse Problem as an example. For each configuration, the ratio of the makespan between the original workflow and its mini-app is $R^{t}_1=0.23$, $R^{t}_2=0.24$, and $R^{t}_3=0.23$; read I/O ratios are $R^{r}_1=0.23$, $R^{r}_2=0.23$, and $R^{r}_3=0.23$; and write I/O ratios are $R^{w}_1=0.23$, $R^{w}_2=0.23$, and $R^{w}_3=0.24$. A constant ratio for each performance metric with all configurations proves that our workflow mini-app performs comparably to the original workflow. 

\begin{table*}[hb]
\centering
\caption{Performance reproducibility for the workflow mini-apps. Performance metrics are collected for each task and the Full workflow. We collected data for both Inverse Problem and DeepDriveMD workflows. We showed a detailed view with the Inverse Problem serial CPU+GPU v1 (first 8 rows) and a summary of DeepDriveMD workflow (last two rows) due to page limit. 
}
\begin{small}
\begin{tabular}{|l|c|c|c|c|c|c|}
\hline
\textbf{Stage} &\textbf{Time(s)} &\textbf{CPU (\%)} &\textbf{GPU (\%)} &\textbf{Read I/O(GB)} &\textbf{Write I/O(GB)} \\
\hline
IP m-app: Simulation Phase 1 & $57.2\pm4.6$ &100 &0  &$39.0\pm0.5$  &$9.7\pm0.0$ \\
\hline
IP m-app: Training Phase 1 &$93.8\pm7.2$ &100 &100 &$12.3\pm0.0$&$0.004\pm0$\\
\hline
IP m-app: Simulation Phase 2 &$54.4\pm3.3$ &100 &0 &$39.7\pm0.0$ &$9.7\pm0.0$\\
\hline
IP m-app: Training Phase 2 &$87.2\pm4.6$ &100 &100 &$12.3\pm0.0$ &$0.004\pm0$\\
\hline
IP m-app: Simulation Phase 3 &$54.6\pm2.3$ &100 &0 &$39.7\pm0.1$ &$9.7\pm0.0$\\
\hline
IP m-app: Training Phase 3 &$84.7\pm0.7$ &100 &100 &$12.3\pm0.0$ &$0.004\pm0$ \\
\hline
IP m-app: Full  &$456.9\pm12.2$ &100 &$58\pm2$&$155.3\pm0.5$ &$29.0\pm0$\\
\hline
IP workflow: Full &$2129.4 \pm 26.0 $ &100 &$62 \pm 1$ &$679.2 \pm 0.3$ & $127.0 \pm 0.1$ \\
\hline
DDMD m-app: Full  &$1436.1\pm10.3$ &$63\pm1$ &$62\pm1$ &$294.8\pm0$ &$148.3\pm0.1$\\
\hline
DDMD workflow: Full &$3054.9\pm 9.2$ &$67\pm1$ &$66\pm1$ &$594.7 \pm 1.1$ &$306.6 \pm 0.4$ \\
\hline
\end{tabular}
\vspace{-10pt}
\label{tab:reproducibility}
\end{small}
\end{table*}

However, looking only at the total makespan, resource and I/O utilization 
can be misleading. Consider the following example: Assume that the total read I/O for both the Inverse Problem workflow and its mini-app is the same; but that, in the original workflow,  
the simulation tasks perform 95\% of the read I/O while in the workflow mini-apps, the training and simulation tasks perform 50\% of the read I/O each. In that scenario, the detailed performance (details similar to Fig.~\ref{fig:originalWF-ru} and Fig.~\ref{fig:originalWF-io}) of the workflow and its mini-app would be very different.
For this reason, we also 
measure resource and I/O utilization during 
the workflow runtime, collecting detailed utilization and makespan data for each of the experiment runs.
Due to the limited space available, we show the CPU+GPU serial execution model from the Inverse Problem and the second configuration of DeepDriveMD (Fig.~\ref{fig:originalWF-ru} and Fig.~\ref{fig:originalWF-io}). That enables us to compare their respective workflow mini-apps with the original workflows. The full results are available in a dedicated GitHub~\cite{mini-apps} repository.

When we compare all the results, we can see that the workflow mini-apps perform similarly to their original workflows in terms of both internal components (tasks) and global execution of the workflow.
For example, the resource utilization pattern of the Inverse Problem workflow V2 (Fig.~\ref{fig:orig_ru_v2}) and its workflow mini-app V2 (Fig.~\ref{fig:wf_mini-app_ru_v2}) are almost identical. Both the I/O size and the execution time share the same fixed ratio (of about 0.23) for all the experiments to reduce the cost of running the workflow mini-apps. Also, for the DeepDriveMD workflow and its workflow mini-app, both resource utilization (Fig.~\ref{fig:orig_ru_ddmd} and Fig.~\ref{fig:wf_mini-app_ru_ddmd}) and I/O (Fig.~\ref{fig:orig_io_ddmd} and Fig.~\ref{fig:wf_mini-app_io_ddmd}) holds the same pattern with a fixed I/O and makespan ratio (of about 0.48). This proves that our workflow mini-app provides performance fidelity both at the workflow level and at the level of its components. Here we remind the reader that the choice of ratio is arbitrary. 
We chose the ratios because of (1) the short makespan of two original workflows (less than an hour) and (2) the makespan of the individual tasks within the workflow. If the original workflow takes longer to execute, we can choose a smaller ratio.

\begin{figure*}[hb]
\centering
    \subfloat[]{\includegraphics[width=0.45\textwidth]{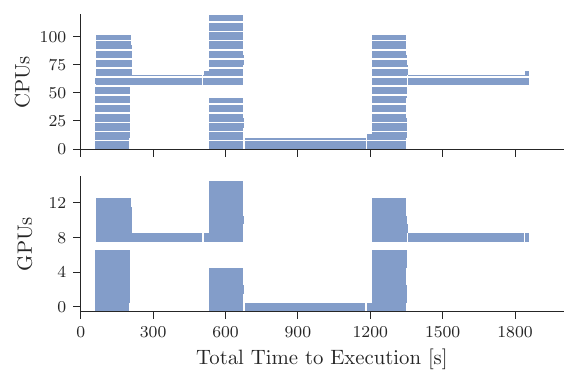}\label{fig:ddmd_frontier_orig}}\hfill
    \subfloat[]{\includegraphics[width=0.45\textwidth]{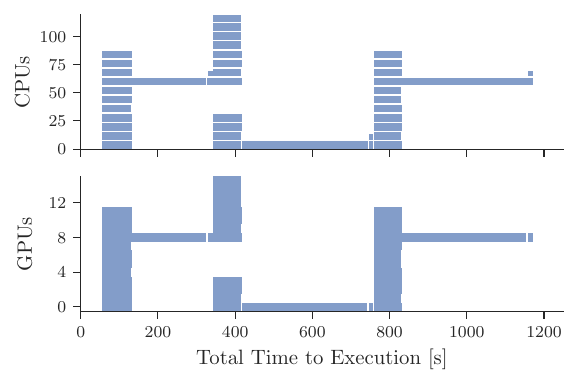}\label{fig:ddmd_frontier_mini}}\hfill
    \\
\caption{\chl{Portability validation of workflow mini-apps. Figures show the CPU and GPU utilization and performance characteristics of the DeepDriveMD original workflow (left) and workflow mini-app (right) on Frontier.}}
\label{fig:portabilityDDMD}
\end{figure*}

\subsection{Performance Reproducibility}
\label{sec:eval_reproducibility}

\begin{figure*}[ht]
\centering
    \subfloat[\centering{RU of workflow mini-app  (CPU+GPU serial v1) \ on Polaris}]{\includegraphics[width=0.33\textwidth]{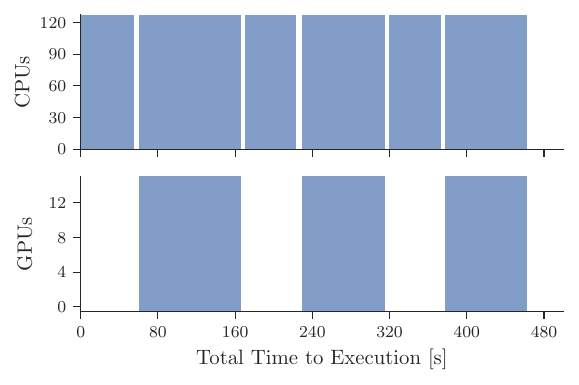}\label{fig:polaris_ru}}\hfill
    \subfloat[\centering{RU of workflow mini-app  (CPU+GPU serial v1) \ on Summit}]{\includegraphics[width=0.33\textwidth]{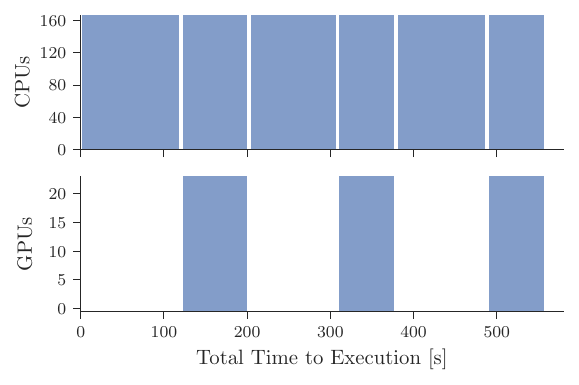}\label{fig:summit_ru}}\hfill
    \subfloat[\centering{RU of workflow mini-app  (CPU+GPU serial v1) \ on Frontier}]{\includegraphics[width=0.33\textwidth]{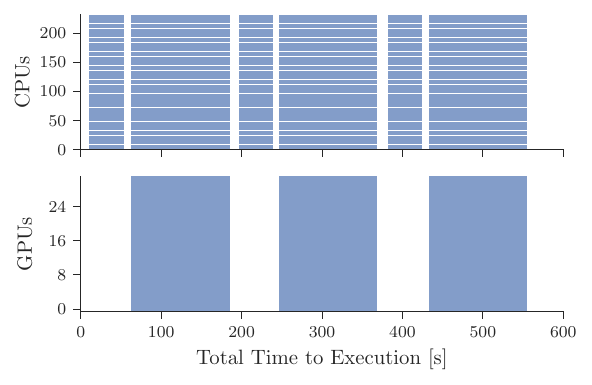}\label{fig:frontier_ru}}\hfill
    \\
    \subfloat[\centering{I/O of workflow mini-app (CPU+GPU serial v1) \ on Polaris}]{\includegraphics[width=0.33\textwidth]{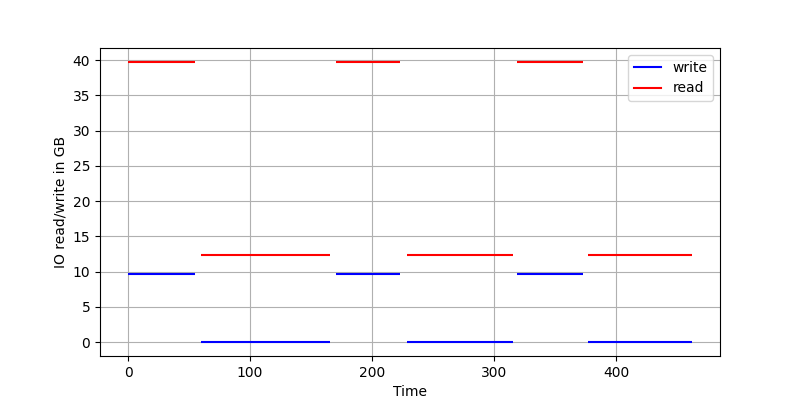}\label{fig:polaris_io}}
    \subfloat[\centering{I/O of workflow mini-app (CPU+GPU serial v1) \ on Summit}]{\includegraphics[width=0.33\textwidth]{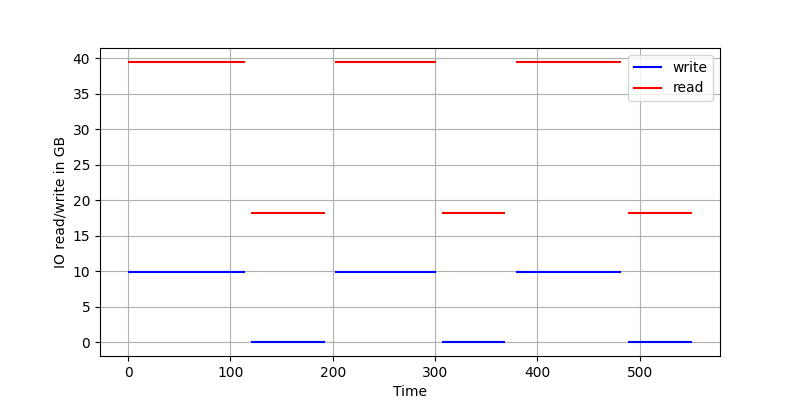}\label{fig:summit_io}}
    \subfloat[\centering{I/O of workflow mini-app (CPU+GPU serial v1) \ on Frontier}]{\includegraphics[width=0.33\textwidth]{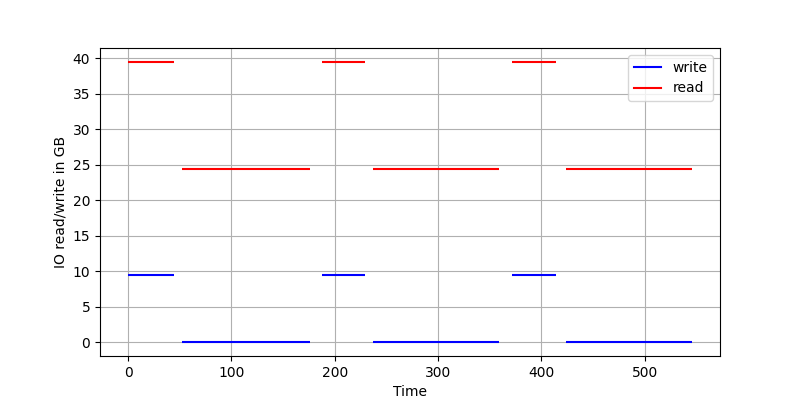}\label{fig:frontier_io}}
\caption{
The performance of the Inverse Problem workflow mini-app on different HPC clusters with an execution model of CPU+GPU serial and experiment configuration V1 as an example. The first three figures show the CPU and GPU resource utilization (RU) as a function of execution time, and the later three figures show the total read/write size of each task during their runtime.}
\vspace{-10pt}
\label{fig:portability}
\end{figure*}

To show that workflow mini-apps can be used for performance reproducibility while requiring less resource and time, we run workflow mini-apps and the original workflows eight times on the same platform (ALCF Polaris) with the same environment. Table~\ref{tab:reproducibility} shows the performance variations between runs. 
Our results show that the Inverse Problem workflow mini-apps had a small performance variation in terms of the makespan of the workflow $<3\%$, which is comparable with the variation of the original workflow of $1.2\%$. A more detailed examination shows that each stage was successfully executed in the intended order, and the performance variation of each stage was $<8\%$. When we focus on the I/O and CPU/GPU utilization, we can see that each task/stage reads and writes almost the same amount of data and utilizes the same resources in each execution. Similarly, the makespan and I/O performance variation for DeepDriveMD and its workflow mini-app is $<1\%$. Thus, combining all metrics (I/O, resource utilization, and makespan) demonstrates that the workflow mini-apps can be used to test and research performance reproducibility of the original workflow with a smaller cost (In this example, cost decreases about a factor of 4 for Inverse Problem and 2 for DeepDriveMD).

\subsection{\chl{Portability}}
\label{sec:eval_portability}

\chl{HPC facilities keep improving. OLCF's Frontier recently became online, and  ALCF's Aurora is next in the line. Testing different platforms is essential to understand how the workflow could perform on them. However, testing each platform with the original workflow increases cost and generally requires code modifications due to complicated dependencies of scientific workflows such as architecture or libraries. }

\chl{For example, the Inverse Problem workflow relies on the GSAS-II library, which can only be run on x86 CPU. The DeepDriveMD workflow relies on openmm, whose deployment on AMD GPU is not trivial, and rapidsai library, which currently only supports NVIDIA GPU. Those dependencies make it difficult to port a workflow from one architecture to another. In this work, for testing and comparison purposes, we try to port the original workflow of DeepDriveMD to the Frontier machine. To do that, we re-implemented the ML task in DeepDriveMD with Tensorflow to get rid of rapidsai dependency so that it could run on AMD GPU, and we spent some effort to deploy everything on Frontier. This takes us around two weeks to finish.}

\chl{Unlike the original workflows, our workflow mini-apps are not dependent on specific libraries and hardware other than generic Python libraries such as NumPy and CuPy that are used to implement wfMiniAPI. The workflow mini-app can be used to test the performance characteristics of the workflow in a new architecture, thanks to its fidelity and portability. It is important to note that performance characteristics focus on workflow execution patterns, bottlenecks, and resource utilization of the workflow. However, similar to any mini-apps, our workflow mini-apps are not built to estimate the exact execution time in a new platform.  
}

We run the workflow mini-apps on three leadership computing facilities (Polaris, Summit, and Frontier) to show both the portability of our workflow mini-apps and analyze how their performance changes in different architectures. We chose these three machines because they all have different architectures. Polaris has AMD EPYC Milan 7543P CPUs and NVIDIA A100 GPUs; Frontier has AMD “Optimized 3rd Gen EPYC” CPUs and AMD Instinct MI250X GPUs; and Summit has IBM POWER9 CPUs and NVIDIA V100 GPUs. 

\chl{For the DeepDriveMD workflow mini-app, we used the workflow mini-app built on Polaris and ported it to Frontier by keeping the number of tasks for each stage the same (12 MD simulations, 1 ML, 1 Selection, and 1 Agent task for each iteration). We changed the number of CPUs assigned to each task based on available resources and scaled the data size to one-quarter. We also had to modify the original DeepDriveMD workflow, as mentioned in the beginning, and run it on Frontier. }

Fig.~\ref{fig:portabilityDDMD} \chl{shows the performance characteristics of both workflow mini-app and modified original workflow. Our results show that we can estimate the resource utilization and performance characteristics correctly by porting workflow mini-app built on a different platform. This provides researchers with what to expect in terms of bottlenecks and resource utilization and helps them to decide whether it is worth modifying the code to use a new platform. }

For the Inverse Problem workflow mini-app (Fig.~\ref{fig:portability}), we again used the workflow mini-app built on Polaris, where we used four nodes with the same setup for each platform; however, nodes in each platform have a different amount of resources (CPU/GPU). We adjust the number of CPUs and GPUs to achieve maximum utilization from each node. We can see that the workflow mini-app mainly holds its execution pattern and I/O; however, performance characteristics and bottlenecks change for each platform. 
Similar to DeepDriveMD, researchers can use this information to make a decision about porting their workflows or not.

\subsection{\chl{Scalability}}
\label{sec:eval_scalability}

\chl{In HPC applications and workflows, scalability (strong scaling) is usually one of the most critical attributes. 
However, it is costly to evaluate the scaling behavior using the original workflow. In this section, we validate the usefulness of the workflow mini-app by comparing its scaling behavior to the original workflow.}

\chl{We picked the Inverse Problem workflow as an example, and the DeepDriveMD workflow gives a similar result. Here, we choose the CPU+GPU serial execution model, and change the number of the nodes, hence the number of CPUs and GPUs, by doubling them while fixing the problem size. We run both the original workflow and the workflow mini-app to get the total makespan and I/O size as a function of the number of nodes.}

\begin{figure*}[htbp]
\centering
  \begin{subfigure}{0.32\textwidth}
      \centering
      \includegraphics[width=1\textwidth]{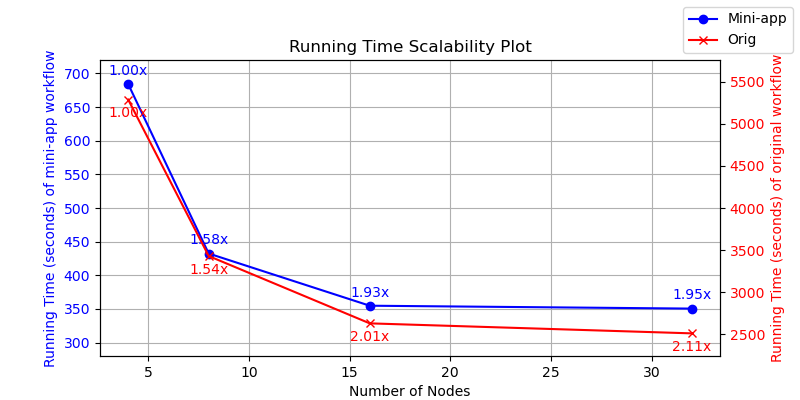}
      \caption{Running time scalability plot.}
      \label{fig:makespan_scalability}
  \end{subfigure}
  \begin{subfigure}{0.32\textwidth}
      \centering
      \includegraphics[width=1\textwidth]{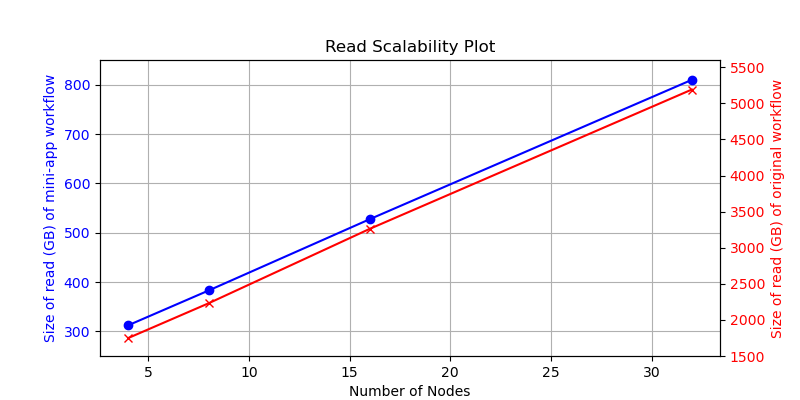}
      \caption{Read scalability plot.}
      \label{fig:read_scalability}
  \end{subfigure}
  \begin{subfigure}{0.32\textwidth}
    \centering
    \includegraphics[width=1\textwidth]{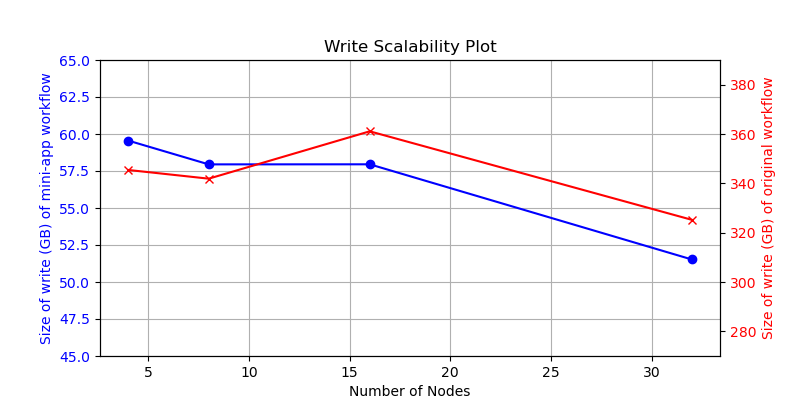}
    \caption{Write scalability plot.}
    \label{fig:write_scalability}
  \end{subfigure}
\caption{\chl{The strong scaling behavior of running time and I/O of the workflow mini-app and the original workflow. The three metrics are collected with 4, 8, 16, and 32 nodes. The factors in subplot (a) are the ratio between running time with node k and the base case (with node 4).}}
\label{fig:scalability}
\vspace{-15pt}
\end{figure*}

\begin{figure*}[htbp]
\centering
  \begin{subfigure}{.45\textwidth}
    \centering
    \includegraphics[width=\linewidth]{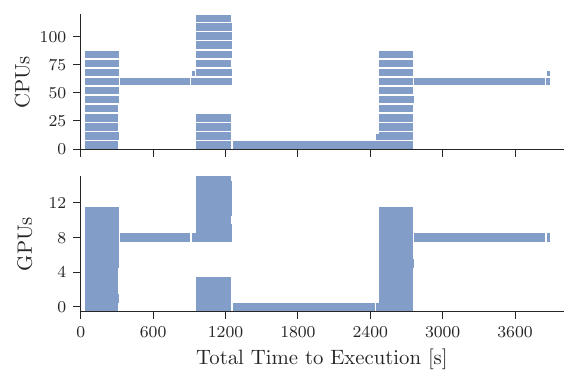}
    \caption{RU of Synchronous}
    \label{fig:ru-ddmd-f-frontier}
  \end{subfigure}%
  \begin{subfigure}{.45\textwidth}
    \centering
    \includegraphics[width=\linewidth]{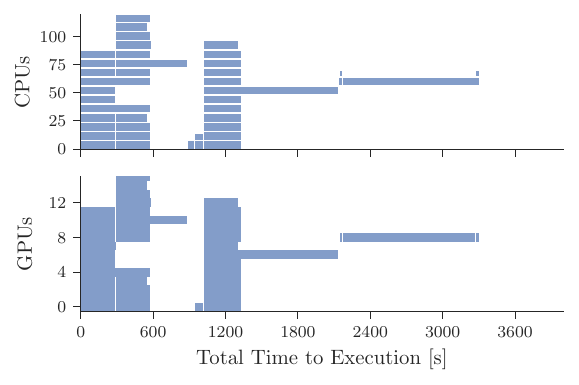}
    \caption{RU of Asynchronious}
    \label{fig:ru-ddmd-async-frontier}
  \end{subfigure}
  \hfill
  \begin{subfigure}{.45\textwidth}
      \centering
      \includegraphics[width=1\textwidth]{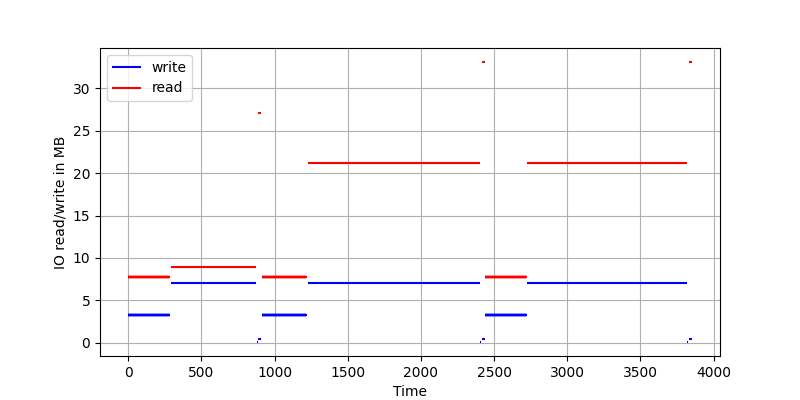}
      \label{fig:io-ddmd-f-frontier}
      \caption{I/O of Synchronous}
  \end{subfigure}
  \begin{subfigure}{.45\textwidth}
      \centering
      \includegraphics[width=1\textwidth]{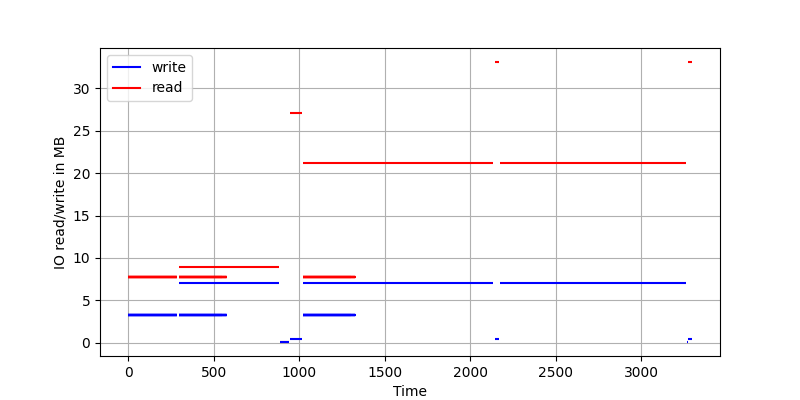}
      \caption{I/O of Asynchronious}
      \label{fig:io-ddmd-asnc-frontier}
  \end{subfigure}
  \caption{The performance of DeepDriveMD workflow mini-app with synchronous and asynchronous execution models. The top two figures show the CPU and GPU resource utilization (RU) as a function of execution time, and the bottom two figures show the total read/write size of each task during their runtime. This figure shows that asynchronous execution of the DeepDriveMD can reduce the makespan by around $10\%$ for this specific case. It also implies that optimizing the training task could benefit mostly the asynchronous execution workflow.}
  \label{fig:ddmd_async}
\hfill
\vspace{-10pt}
\end{figure*}

Fig.~\ref{fig:scalability} \chl{shows the performance of the workflow mini-app versus the original workflow using 4, 8, 16, and 32 nodes, respectively. We can see that the scaling behavior of the original workflow and workflow mini-app looks similar, suggesting that our workflow mini-app could be used to measure the scaling behavior of the original workflow with high fidelity.} As shown in Fig.~\ref{fig:makespan_scalability}, \chl{the makespan of both the workflows improves by roughly a factor of 1.5 from 4 nodes to 8 nodes, but the performance improvement slows down afterward. The makespan almost flattens after 16 nodes. If we look at the initial discussion of the performance of the original workflow}~\cite{wang2023parallel}, \chl{the authors did a scaling test for each task separately and found that the ML tasks did not have a good strong scaling behavior, which is consistent with what we found here. }

\chl{When we look at the effect of the scaling on the I/O} (Fig.~\ref{fig:read_scalability} and Fig.~\ref{fig:write_scalability}), \chl{we can see that while the write I/O size remains stable with the increased number of nodes, read I/O increases linearly concerning the number of the nodes. This is because each rank of training task would need to read the entire dataset created from the simulation task, which increases the read I/O on par with the number of training tasks. In summary, these suggest our workflow mini-app could be helpful for the scientists to not only measure but also study the scaling of the original workflow to get the maximum performance at a smaller cost.}

\subsection{Testing Execution Models}
\label{sec:executionmodel}

To test a new execution model, we implemented an asynchronous version of the DeepDriveMD workflow mini-app, which is built based on Ref.~\cite{pascuzziasynchronous}. We implement this by overlapping training and the next set of MD simulations.  We run both synchronous and asynchronous versions of the DeepDriveMD workflow mini-apps on Frontier. Fig.~\ref{fig:ddmd_async} shows how makespan, resource utilization, and I/O change.  We can see that with asynchronous execution, we were able to increase both CPU and GPU utilization by running MD simulation\_i and training\_i+1 simultaneously. This by itself improves the performance of the workflow mini-app by 10\%. This detailed information is helpful in understanding if it is worth the time to modify the workflow to generate a new execution model and, if we decide to choose this execution model, which task should be optimized to improve the overall performance. (For example, the overlapping regions in Fig.~\ref{fig:ru-ddmd-async-frontier} imply that optimizing the training task contributes most to the overall performance). Those can be collected with minimal cost (implementation effort and execution cost).

\section{Limitations}
\label{sec:limitations}

\chl{One limitation of workflow mini-apps is their building process. Specifically,  developers need to be familiar with the tasks of the actual workflow to build a high-fidelity emulated version of those tasks. Usually, emulators can be built by profiling the initial task several times, but sometimes domain knowledge about the initial task is still needed due to the intrinsic limitations of a profiler. Automating or at least assisting in the building of emulated high-fidelity versions of general-purpose tasks is one of the ongoing research topics of this project. 
Nonetheless, while emulating workflow tasks is a nontrivial process, it enables a great reduction of the time and resources needed to study the portability, behavior, and performance of the actual workflow. }

\chl{Workflow mini-apps can be used to predict the performance characteristics of their corresponding workflows with a new configuration, execution model, or on a new system. However, mini-apps cannot be used to predict performance changes caused by scientific input changes, especially if they are triggering scientific calculations-based conditionals. Capturing the performance based on scientific input changes requires running actual science which will negate two of the main ideas of mini-apps, i.e., simplicity and efficiency. However, we are working alongside the workflow community to extend the wfMiniAPI to cover more corner cases. 
}

\section{Future Works}
\label{sec:futureworks}

\chl{We are working on reducing the need for domain knowledge and making our workflow mini-apps more generic. For this purpose, we are working on automating task emulation by (1) integrating additional performance monitoring tools that could provide a larger amount of performance data and (2) extending the mini-apps API to include more diverse kernels that can mimic the performance of critical bottlenecks. Together, granular performance data and the API extensions will help us to achieve our goal to automate the creation of emulated tasks. }

\chl{Our workflow mini-app can represent any type of workflow. For example, we are working on generating workflow mini-apps for a complex high-energy physics workflow} (DUNE~\cite{abi2020volume}) \chl{with more than 16 different types of tasks and complicated dependencies. To generalize workflow mini-apps even further, we are looking to integrate more options in the workflow generation step, such as a Directed Acyclic Graph (DAG) analyzer to assist in building the workflow mini-apps using emulated tasks.}  

\chl{Finally, 
we also made our API public and are encouraging our community to contribute to the wfMiniAPI. With this, we are aiming to provide a more modular framework.}

\section{Conclusion}
\label{sec:conclusion}

In this paper, we proposed workflow mini-apps for addressing the challenges of building, testing, and optimizing scientific workflows. We provide a methodology to design workflow mini-apps and show our implementation for real-world scientific workflows. Our evaluation shows that the workflow mini-app can provide simplicity and portability while keeping the performance fidelity to the original workflow. On top of that, workflow mini-apps can be used to study the performance reproducibility of scientific workflows. We show that workflow mini-apps can be deployed and run on various HPC systems (Summit, Frontier, and Polaris), CPU architectures (PowerPC and Intel x86),  and GPU  architectures (NVIDIA and AMD) without workflow-specific constraints.

We also introduced our workflow mini-apps library API to reduce the difficulty of building emulated tasks. Our API is open source, and we encourage community involvement to extend its capabilities.

\chl{Finally, we mentioned our current limitations and how we are planning to address them in our future works.}

\section*{Acknowledgements}
\footnotesize{This work was supported by DOE ASCR 0269227 and DEAC02-06CH11357 (RECUP), ECP ExaWorks, and DOE HEP B\&R KA2401045 Center for Computational Excellence at BNL, as well as NSF-1931512 (RADICAL-Cybertools). This research used resources at ALCF ANL and OLCF ORNL, which are supported by the Office of Science of the U.S. Department of Energy under Contract DE-AC02-06CH11357 and DE-AC05-00OR22725 respectively.}
\vspace{25pt}

\vspace{-5pt}
\bibliographystyle{IEEEtran}
\bibliography{main,radical}

\begin{thebibliography}{10}
\providecommand{\url}[1]{#1}
\csname url@samestyle\endcsname
\providecommand{\newblock}{\relax}
\providecommand{\bibinfo}[2]{#2}
\providecommand{\BIBentrySTDinterwordspacing}{\spaceskip=0pt\relax}
\providecommand{\BIBentryALTinterwordstretchfactor}{4}
\providecommand{\BIBentryALTinterwordspacing}{\spaceskip=\fontdimen2\font plus
\BIBentryALTinterwordstretchfactor\fontdimen3\font minus
  \fontdimen4\font\relax}
\providecommand{\BIBforeignlanguage}[2]{{%
\expandafter\ifx\csname l@#1\endcsname\relax
\typeout{** WARNING: IEEEtran.bst: No hyphenation pattern has been}%
\typeout{** loaded for the language `#1'. Using the pattern for}%
\typeout{** the default language instead.}%
\else
\language=\csname l@#1\endcsname
\fi
#2}}
\providecommand{\BIBdecl}{\relax}
\BIBdecl

\bibitem{casalino2020aidriven}
L.~Casalino, A.~Dommer, Z.~Gaieb \emph{et~al.}, ``Ai-driven multiscale
  simulations illuminate mechanisms of sars-cov-2 spike dynamics,'' 2020.

\bibitem{covidisairborne2021ijhpca}
A.~Dommer, L.~Casalino, F.~Kearns \emph{et~al.}, ``$\#$covidisairborne:
  Ai-enabled multiscale computational microscopy of delta sars-cov-2 in a
  respiratory aerosol,'' \emph{International Journal of High-Performance
  Computing Applications}, 2021.

\bibitem{hepccereport2023indigo}
\BIBentryALTinterwordspacing
C.~Leggett, P.~Calafiura, O.~Gutsche \emph{et~al.} (2023, May) 26th
  international conference on computing in high energy \& nuclear physics
  (chep2023). [Online]. Available:
  \url{https://indico.jlab.org/event/459/contributions/11821/}
\BIBentrySTDinterwordspacing

\bibitem{hepccerwebpage}
\BIBentryALTinterwordspacing
``Hep-cce webpage.'' [Online]. Available:
  \url{https://www.anl.gov/hep-cce/complex-workflows}
\BIBentrySTDinterwordspacing

\bibitem{aad2008atlas}
G.~Aad, X.~S. Anduaga, S.~Antonelli \emph{et~al.}, ``The atlas experiment at
  the cern large hadron collider,'' 2008.

\bibitem{abi2020volume}
B.~Abi, R.~Acciarri, M.~A. Acero \emph{et~al.}, ``Volume i. introduction to
  dune,'' \emph{Journal of instrumentation}, vol.~15, no.~08, p. T08008, 2020.

\bibitem{fogerty17}
S.~Fogerty, S.~Bishnu, Y.~Zamora \emph{et~al.}, ``Thoughtful precision in
  mini-apps,'' in \emph{2017 IEEE International Conference on Cluster Computing
  (CLUSTER)}, 2017, pp. 858--865.

\bibitem{sukumar16}
S.~R. Sukumar, M.~A. Matheson, R.~Kannan, and S.-H. Lim, ``Mini-apps for high
  performance data analysis,'' in \emph{2016 IEEE International Conference on
  Big Data (Big Data)}, 2016, pp. 1483--1492.

\bibitem{wang2023parallel}
T.~Wang, S.~K. Seal, R.~Kannan, C.~Garcia-Cardona, T.~Proffen, and S.~Jha, ``A
  parallel machine learning workflow for neutron scattering data analysis,'' in
  \emph{2023 IEEE International Parallel and Distributed Processing Symposium
  Workshops (IPDPSW)}.\hskip 1em plus 0.5em minus 0.4em\relax IEEE, 2023, pp.
  795--798.

\bibitem{brace2022coupling}
A.~Brace, I.~Yakushin, H.~Ma, A.~Trifan, T.~Munson, I.~Foster, A.~Ramanathan,
  H.~Lee, M.~Turilli, and S.~Jha, ``Coupling streaming ai and hpc ensembles to
  achieve 100--1000$\times$ faster biomolecular simulations,'' in \emph{2022
  IEEE International Parallel and Distributed Processing Symposium
  (IPDPS)}.\hskip 1em plus 0.5em minus 0.4em\relax IEEE, 2022, pp. 806--816.

\bibitem{nicolae2023building}
B.~Nicolae, T.~Z. Islam, R.~Ross \emph{et~al.}, ``Building the i
  (interoperability) of fair for performance reproducibility of large-scale
  composable workflows in recup,'' in \emph{2023 IEEE 19th International
  Conference on e-Science (e-Science)}.\hskip 1em plus 0.5em minus 0.4em\relax
  IEEE, 2023, pp. 1--7.

\bibitem{Patki_Thiagarajan_Ayala_Islam_2019}
T.~Patki, J.~J. Thiagarajan, A.~Ayala, and T.~Z. Islam, ``Performance
  optimality or reproducibility: that is the question,'' in \emph{Proceedings
  of the International Conference for High Performance Computing, Networking,
  Storage and Analysis}, ser. SC ’19.\hskip 1em plus 0.5em minus 0.4em\relax
  New York, NY, USA: Association for Computing Machinery, Nov 2019, p. 1–30.

\bibitem{ecpProxy}
\BIBentryALTinterwordspacing
``Ecp proxy applications,'' accessed on August 13, 2023. [Online]. Available:
  \url{https://proxyapps.exascaleproject.org}
\BIBentrySTDinterwordspacing

\bibitem{matsuoka2022preparing}
S.~Matsuoka, J.~Domke, M.~Wahib \emph{et~al.}, ``Preparing for the
  future—rethinking proxy applications,'' \emph{Computing in Science \&
  Engineering}, vol.~24, no.~2, pp. 85--90, 2022.

\bibitem{wu2019performance}
X.~Wu, V.~Taylor, J.~M. Wozniak, R.~Stevens, T.~Brettin, and F.~Xia,
  ``Performance, energy, and scalability analysis and improvement of parallel
  cancer deep learning candle benchmarks,'' in \emph{Proceedings of the 48th
  International Conference on Parallel Processing}, 2019, pp. 1--11.

\bibitem{thompson2018ecp}
A.~Thompson, A.~Cangi, C.~Trott, C.~Junghans, S.~Moore, and T.~Germann,
  ``Ecp-copa/examinimd,'' Sandia National Lab.(SNL-NM), Albuquerque, NM (United
  States), Tech. Rep., 2018.

\bibitem{ghosh2018minivite}
S.~Ghosh, M.~Halappanavar, A.~Tumeo, A.~Kalyanaraman, and A.~H. Gebremedhin,
  ``minivite: A graph analytics benchmarking tool for massively parallel
  systems,'' in \emph{2018 IEEE/ACM Performance Modeling, Benchmarking and
  Simulation of High Performance Computer Systems (PMBS)}.\hskip 1em plus 0.5em
  minus 0.4em\relax IEEE, 2018, pp. 51--56.

\bibitem{nersc}
\BIBentryALTinterwordspacing
``Nersc-10 benchmarks,'' accessed on August 11, 2023. [Online]. Available:
  \url{https://www.nersc.gov/systems/nersc-10/benchmarks/}
\BIBentrySTDinterwordspacing

\bibitem{gadioli2021tunable}
D.~Gadioli, G.~Palermo, S.~Cherubin \emph{et~al.}, ``Tunable approximations to
  control time-to-solution in an hpc molecular docking mini-app,'' \emph{The
  Journal of Supercomputing}, vol.~77, pp. 841--869, 2021.

\bibitem{vineyard2022neural}
C.~Vineyard, S.~Cardwell, F.~Chance \emph{et~al.}, ``Neural mini-apps as a tool
  for neuromorphic computing insight,'' in \emph{Proceedings of the 2022 Annual
  Neuro-Inspired Computational Elements Conference}, 2022, pp. 40--49.

\bibitem{martineau2017arch}
M.~Martineau and S.~McIntosh-Smith, ``The arch project: physics mini-apps for
  algorithmic exploration and evaluating programming environments on hpc
  architectures,'' in \emph{2017 IEEE International Conference on Cluster
  Computing (CLUSTER)}.\hskip 1em plus 0.5em minus 0.4em\relax IEEE, 2017, pp.
  850--857.

\bibitem{ewart2017neuromapp}
T.~Ewart, J.~Planas, F.~Cremonesi, K.~Langen, F.~Sch{\"u}rmann, and
  F.~Delalondre, ``Neuromapp: a mini-application framework to improve neural
  simulators,'' in \emph{High Performance Computing: 32nd International
  Conference, ISC High Performance 2017, Frankfurt, Germany, June 18--22, 2017,
  Proceedings 32}.\hskip 1em plus 0.5em minus 0.4em\relax Springer, 2017, pp.
  181--198.

\bibitem{ramesh2022ghost}
S.~Ramesh, M.~Titov, M.~Turilli, S.~Jha, and A.~Malony, ``The ghost of
  performance reproducibility past,'' in \emph{2022 IEEE 18th International
  Conference on e-Science (e-Science)}.\hskip 1em plus 0.5em minus 0.4em\relax
  IEEE, 2022, pp. 513--518.

\bibitem{pouchard2019computational}
L.~Pouchard, S.~Baldwin, T.~Elsethagen, S.~Jha, B.~Raju, E.~Stephan, L.~Tang,
  and K.~K. Van~Dam, ``Computational reproducibility of scientific workflows at
  extreme scales,'' \emph{The Int. Journal of High Performance Computing
  Applications}, vol.~33, no.~5, pp. 763--776, 2019.

\bibitem{pathway2013}
V.~Petkov, M.~Gerndt, and M.~Firbach, ``Pathway: Performance analysis and
  tuning using workflows,'' in \emph{2013 IEEE 10th International Conference on
  High Performance Computing and Communications \& 2013 IEEE International
  Conference on Embedded and Ubiquitous Computing}, 2013, pp. 792--799.

\bibitem{benedict2013performance}
S.~Benedict, ``Performance issues and performance analysis tools for hpc cloud
  applications: a survey,'' \emph{Computing}, vol.~95, no.~2, pp. 89--108,
  2013.

\bibitem{luttgau2018toward}
J.~L{\"u}ttgau, S.~Snyder, P.~Carns, J.~M. Wozniak, J.~Kunkel, and T.~Ludwig,
  ``Toward understanding i/o behavior in hpc workflows,'' in \emph{2018
  IEEE/ACM 3rd international workshop on parallel data storage \& data
  intensive scalable computing systems (PDSW-DISCS)}.\hskip 1em plus 0.5em
  minus 0.4em\relax IEEE, 2018, pp. 64--75.

\bibitem{mattoso2015dynamic}
M.~Mattoso, J.~Dias, K.~A. Ocana \emph{et~al.}, ``Dynamic steering of hpc
  scientific workflows: A survey,'' \emph{Future Generation Computer Systems},
  vol.~46, pp. 100--113, 2015.

\bibitem{kilic2022memgaze}
O.~O. Kilic, N.~R. Tallent, Y.~Suriyakumar, C.~Xie, A.~Marquez, and S.~Eranian,
  ``Memgaze: Rapid and effective load-level memory trace analysis,'' in
  \emph{2022 IEEE International Conference on Cluster Computing
  (CLUSTER)}.\hskip 1em plus 0.5em minus 0.4em\relax IEEE, 2022, pp. 484--495.

\bibitem{kilic2020rapid}
O.~O. Kilic, N.~R. Tallent, and R.~D. Friese, ``Rapid memory footprint access
  diagnostics,'' in \emph{2020 IEEE International Symposium on Performance
  Analysis of Systems and Software (ISPASS)}.\hskip 1em plus 0.5em minus
  0.4em\relax IEEE, 2020, pp. 273--284.

\bibitem{kelly2020chimbuko}
C.~Kelly, S.~Ha, K.~Huck, H.~Van~Dam, L.~Pouchard, G.~Matyasfalvi, L.~Tang,
  N.~D'Imperio, W.~Xu, S.~Yoo \emph{et~al.}, ``Chimbuko: A workflow-level
  scalable performance trace analysis tool,'' in \emph{ISAV'20 In Situ
  Infrastructures for Enabling Extreme-Scale Analysis and Visualization}, 2020,
  pp. 15--19.

\bibitem{mini-apps}
\BIBentryALTinterwordspacing
``Workflow mini-apps github repo:.'' [Online]. Available:
  \url{https://github.com/radical-cybertools/workflow-mini-apps/}
\BIBentrySTDinterwordspacing

\bibitem{balasubramanian2019radical}
\BIBentryALTinterwordspacing
V.~Balasubramanian, S.~Jha, A.~Merzky, and M.~Turilli, ``Radical-cybertools:
  Middleware building blocks for scalable science,'' \emph{CoRR}, vol.
  abs/1904.03085, 2019. [Online]. Available:
  \url{http://arxiv.org/abs/1904.03085}
\BIBentrySTDinterwordspacing

\bibitem{entk2016bala}
V.~Balasubramanian, A.~Treikalis, O.~Weidner, and S.~Jha, ``Ensemble toolkit:
  Scalable and flexible execution of ensembles of tasks,'' in \emph{2016 45th
  International Conference on Parallel Processing (ICPP)}, vol.~00, Aug. 2016,
  pp. 458--463.

\bibitem{maeno2011overview}
T.~Maeno, K.~De, T.~Wenaus \emph{et~al.}, ``Overview of atlas panda workload
  management,'' in \emph{Journal of Physics: Conference Series}, vol. 331,
  no.~7.\hskip 1em plus 0.5em minus 0.4em\relax IOP Publishing, 2011, p.
  072024.

\bibitem{ward2021colmena}
L.~Ward, G.~Sivaraman, J.~G. Pauloski \emph{et~al.}, ``Colmena: Scalable
  machine-learning-based steering of ensemble simulations for high performance
  computing,'' in \emph{2021 IEEE/ACM Workshop on Machine Learning in High
  Performance Computing Environments (MLHPC)}.\hskip 1em plus 0.5em minus
  0.4em\relax IEEE, 2021, pp. 9--20.

\bibitem{alexander2021co}
F.~J. Alexander, J.~Ang, J.~A. Bilbrey, J.~Balewski, T.~Casey, R.~Chard,
  J.~Choi, S.~Choudhury, B.~Debusschere, A.~M. DeGennaro \emph{et~al.},
  ``Co-design center for exascale machine learning technologies (exalearn),''
  \emph{The International Journal of High Performance Computing Applications},
  vol.~35, no.~6, pp. 598--616, 2021.

\bibitem{lee2019deepdrivemd}
H.~Lee, M.~Turilli, S.~Jha, D.~Bhowmik, H.~Ma, and A.~Ramanathan,
  ``Deepdrivemd: Deep-learning driven adaptive molecular simulations for
  protein folding,'' in \emph{2019 IEEE/ACM Third Workshop on Deep Learning on
  Supercomputers (DLS)}.\hskip 1em plus 0.5em minus 0.4em\relax IEEE, 2019, pp.
  12--19.

\bibitem{radical_analytics_url}
``{RADICAL-Analytics Github Project},'' \newline
  \url{https://github.com/radical-cybertools/radical.analytics}.

\bibitem{carns200924}
P.~Carns, R.~Latham, R.~Ross \emph{et~al.}, ``24/7 characterization of
  petascale i/o workloads,'' in \emph{2009 IEEE International Conference on
  Cluster Computing and Workshops}.\hskip 1em plus 0.5em minus 0.4em\relax
  IEEE, 2009, pp. 1--10.

\bibitem{sarisky2001betabetaalpha}
C.~A. Sarisky and S.~L. Mayo, ``The $\beta$$\beta$$\alpha$ fold: explorations
  in sequence space,'' \emph{Journal of molecular biology}, vol. 307, no.~5,
  pp. 1411--1418, 2001.

\bibitem{pascuzziasynchronous}
V.~R. Pascuzzi, O.~O. Kilic, M.~Turilli, and S.~Jha, ``Asynchronous execution
  of heterogeneous tasks in ml-driven hpc workflows,'' \emph{Job Scheduling
  Strategies for Parallel Processing: 26th International Workshop, JSSPP
  2023,}, 2023.

\end{thebibliography}




\end{document}